\documentclass[aps, prl, superscriptaddress,amsmath,amssymb,twocolumn,longbibliography]{revtex4-2}
\usepackage{graphicx}
\usepackage{subfigure}
\usepackage{adjustbox}
\usepackage{bm}
\usepackage{color}
\usepackage{braket}
\usepackage{standalone}
\usepackage{multirow}
\usepackage{tikz}
\usepackage{mathrsfs}
\usepackage{dsfont}
\usepackage{comment}
\usepackage{amsmath}
\usepackage[colorlinks,bookmarks=true,citecolor=blue,linkcolor=red,urlcolor=blue]{hyperref}
\usepackage{cleveref}
\usepackage{orcidlink}
\usepackage{float}
\usepackage{xcolor}
\usepackage{pdfpages}
\makeatletter
\AtBeginDocument{\let\LS@rot\@undefined}
\makeatother

\newcommand{\bea}{\begin{eqnarray*}}
\newcommand{\eea}{\end{eqnarray*}}
\newcommand{\bean}{\begin{eqnarray}}
\newcommand{\eean}{\end{eqnarray}}
\newcommand{\lgl}{\langle}
\newcommand{\rgl}{\rangle}
\newcommand{\nn}{\nonumber\\}

\renewcommand{\rm}{\mathrm}
\newcommand{\bs}{{\mathbf s}}
\newcommand{\br}{{\mathbf r}}
\newcommand{\bn}{{\mathbf n}}

\newcommand{\bU}{{\mathbf U}}

\newcommand{\btau}{\bm{\tau}}

\newcommand{\hc}{\hat c}

\newcommand{\hcd}{\hat c^\dagger}

\newcommand{\bwe}{\begin{widetext}\begin{eqnarray}}
\newcommand{\eew}{\end{eqnarray}\end{widetext}}

\definecolor{BOcolor}{RGB}{0, 128, 0}
\definecolor{CBOcolor}{RGB}{0, 255, 255}
\definecolor{SVExcolor}{RGB}{210, 180, 140}
\definecolor{SVEcolor}{RGB}{152, 251, 152}
\definecolor{coexistcolorz}{RGB}{255, 105, 180}
\definecolor{coexistcolorv}{RGB}{255, 0, 255}
\definecolor{coexistcolorzv}{RGB}{255, 165, 0}

\newcommand{\BOcircle}{\tikz{\draw[draw=black, fill=BOcolor] (0,0) circle (0.9ex)}}

\definecolor{twoCSbofm}{RGB}{128, 128, 128}

\definecolor{twoCSboafm}{RGB}{230, 230, 250}

\definecolor{twoCSbocafm}{RGB}{165, 42, 42}

\definecolor{twoCSsve1}{RGB}{152, 251, 152}
\newcommand{\twoCSsveONE}{\tikz{\draw[draw=black, fill=twoCSsve1] (0,0) circle (0.9ex)}}
\definecolor{twoCSsve2}{RGB}{255, 165, 0}
\newcommand{\twoCSsveTWO}{\tikz{\draw[draw=black, fill=twoCSsve2] (0,0) circle (0.9ex)}}
\definecolor{twoCSsve3}{RGB}{75, 0, 130}

\definecolor{twoCSsve4}{RGB}{255, 105, 180}
\newcommand{\twoCSsveFOUR}{\tikz{\draw[draw=black, fill=twoCSsve4] (0,0) circle (0.9ex)}}
\definecolor{twoCSsve5}{RGB}{70, 130, 180}

\definecolor{twoCSsve6}{RGB}{0, 128, 128}

\begin{document}

\title{Anomalous Transport Gaps of Fractional Quantum Hall Phases in Graphene Landau Levels are Induced by Spin-Valley Entangled Ground States}

\author{Jincheng An}
\email{jincheng.an1@gmail.com}
\affiliation{Department of Physics and Astronomy, University of Kentucky, Lexington, KY 40506, USA}
\author{Ajit C. Balram\orcidlink{0000-0002-8087-6015}}
\email{cb.ajit@gmail.com}
\affiliation{Institute of Mathematical Sciences, CIT Campus, Chennai, 600113, India}
\affiliation{Homi Bhabha National Institute, Training School Complex, Anushaktinagar, Mumbai 400094, India}
\author{Udit Khanna\orcidlink{0000-0002-3664-4305}}
\email{udit.khanna.10@gmail.com}
\affiliation{Theoretical Physics Division, Physical Research Laboratory, Navrangpura, Ahmedabad-380009, India}
\author{Ganpathy Murthy\orcidlink{0000-0001-8047-6241}}
\email{murthy@g.uky.edu}
\affiliation{Department of Physics and Astronomy, University of Kentucky, Lexington, KY 40506, USA}	

\begin{abstract}
We evaluate the transport gaps in the most prominent fractional quantum Hall states in the $\bn{=}0$ and $\bn{=}1$ Landau Levels of graphene, accounting for the Coulomb interaction, lattice-scale anisotropies, and one-body terms. We find that the fractional phases in the $\bn{=}0$ Landau level are bond-ordered, while those in the $\bn{=}1$ Landau level are spin-valley entangled. This resolves a long-standing experimental puzzle [Amet, \emph{et al.}, Nat. Comm. {\bf 6}, 5838 (2015)] of the contrasting Zeeman dependence of the transport gaps in the two Landau levels. The spin-valley entangled phases host gapless Goldstone modes that can be probed via bulk thermal transport measurements. As a byproduct of our computations, we place strong constraints on the values of the microscopic anisotropic couplings such that these are consistent with all known experimental results. 
\end{abstract}

\maketitle

{\bf \em Introduction.---}
Fractional quantum Hall (FQH) states are quintessentially strongly correlated topological states with protected chiral edge modes. Discovered in 1982 in two-dimensional electron gases~\cite{FQHE_Discovery_1982}, FQH states have fractionally charged excitations with fractional statistics~\cite{Laughlin_1983, Arovas84}. While FQH states do not need any symmetry-breaking to exist, in the presence of internal degrees of freedom such as spin or valley, spontaneous symmetry breaking driven by electron-electron interactions can occur in the quantum Hall regime, a phenomenon known as quantum Hall ferromagnetism (QHFM)~\cite{QHFM_Fertig_1989, Shivaji_Skyrmion, QHFM_Yang_etal_1994, QHFM_Moon_etal_1995}.

The discovery of graphene~\cite{Novoselov_etal_2004, Novoselov_et_al_2005, Zhang_Kim_2005}, a two-dimensional honeycomb lattice of carbon atoms, opened a new platform for quantum Hall physics. In addition to spin, graphene has two degenerate valleys at the two inequivalent corners ($\mathbf{K}$ and $\mathbf{K}^\prime$) of the Brillouin zone (BZ). The wavefunctions in each valley obey a two-dimensional Dirac equation at low energies. In the presence of a perpendicular magnetic field $B_\perp$, the non-interacting spectrum exhibits nearly four-fold (spin and valley) degenerate Landau levels (LLs) that are particle-hole symmetric, with orbital energies $E_{\pm \bn} {\propto} {\pm}\sqrt{\bn B_\perp}$.
The zeroth LL (ZLL) manifold uniquely features valley-sublattice locking: states in the $\mathbf{K}$ valley reside on the $B$ sublattice, while those in the $\mathbf{K}^\prime$ valley occupy the $A$ sublattice. 

Numerous integer and fractional quantum Hall states have been seen in both the $\bn{=}0$ and higher $\bn$-th LL manifolds of graphene~\cite{Novoselov_et_al_2005, Zhang_Kim_2005, Bolotin_FQH2009, young2012:nu0, Phase_transition_vs_Bperp_Yacoby2013, young2014:nu0, GoldhaberGordon2015, Even_Den_Young2018, Kim19}. The vast majority of these states show QHFM, driven by a combination of the Zeeman field $E_Z$, the sublattice potential $E_V$, the long-range Coulomb interaction (possessing a $SU(4)$ symmetry in the spin/valley space), and residual lattice-scale valley anisotropic interactions. These residual interactions lower the symmetry of the two-body interactions to $SU(2)_{\text{spin}} {\otimes} U(1)_{\text{valley}}$~\cite{alicea2006:gqhe}.  

In a seminal study based on previous works~\cite{abanin2006:nu0,brey2006:nu0,alicea2006:gqhe, Herbut1, Herbut2, KYang_SU4_Skyrmion_2006}, Kharitonov~\cite{kharitonov2012:nu0} analyzed the filling $\nu{=}0$ state (charge neutrality, at which two of the four levels in the ZLL-manifold are full) and found that the ground state may be in one of four phases depending on the anisotropic couplings: ferromagnetic (FM), canted antiferromagnetic (CAFM), bond-ordered (BO), and charge density wave (CDW). Since the lattice scale is much smaller than the typical magnetic length, $\ell {=}\sqrt{\hbar c/(e B_{\perp})}$, Kharitonov assumed an ultra-short-range (USR) form for the anisotropic interactions. If the USR condition is relaxed, additional phases are found at $\nu{=}0$~\cite{Das_Kaul_Murthy_2022, De_etal_Murthy2022, Stefanidis_Sodemann2022} and at fractional $\nu$~\cite{Sodemann_MacDonald_2014, Jincheng2024}. Based on transport~\cite{young2014:nu0}, magnon transmission~\cite{Magnon_transport_Yacoby_2018}, and scanning tunneling microscopy (STM)~\cite{STM_Yazdani2021visualizing, STM_Coissard_2022}, it is believed that the anisotropic couplings are such that the $\nu{=}0$ system is either in the CAF or the BO phase at perpendicular field. 

Transport experiments probing the FQH states exhibit puzzling qualitative differences between states in the ZLLs and higher LL manifolds. Amet {\it et al}~\cite{GoldhaberGordon2015} measured the transport gaps in tilted fields for various fractions in the ZLLs and the $\bn{=}1$ LL manifold (1LLs) of graphene. They observed that the transport gaps for the FQH states in the ZLLs are independent of the Zeeman coupling $E_Z{\propto} B_\text{tot}{=}\sqrt{B^2_\perp{+}B^2_{\parallel}}$ under varying $B_{\parallel}$, but decrease with $E_Z$ in the 1LLs. While FQH states in the ZLLs can be in spin-singlet states~\cite{Sodemann_MacDonald_2014, Balram15a, Jincheng2024}, in the 1LLs they are spontaneously spin-polarized even at $E_Z{=}0$~\cite{Balram15c}, thereby ruling out spin transitions. Consequently, the persistence of a $E_Z$ dependence of the transport gap even at total fields as high as $B_\text{tot}{=} 44$ T has been one of the longstanding puzzles in the field. In contrast, transitions between different spin/valley-polarized ground states in the ZLLs have been observed at many fractions, and these are in good agreement with the theory of free composite fermions~\cite{Jain89, Phase_transition_vs_Bperp_Yacoby2013, Abanin_Halperin2013, Balram15a, HUang21}.

\begin{figure}[H]
    \centering \includegraphics[width=0.49\textwidth,height=0.13\textwidth]{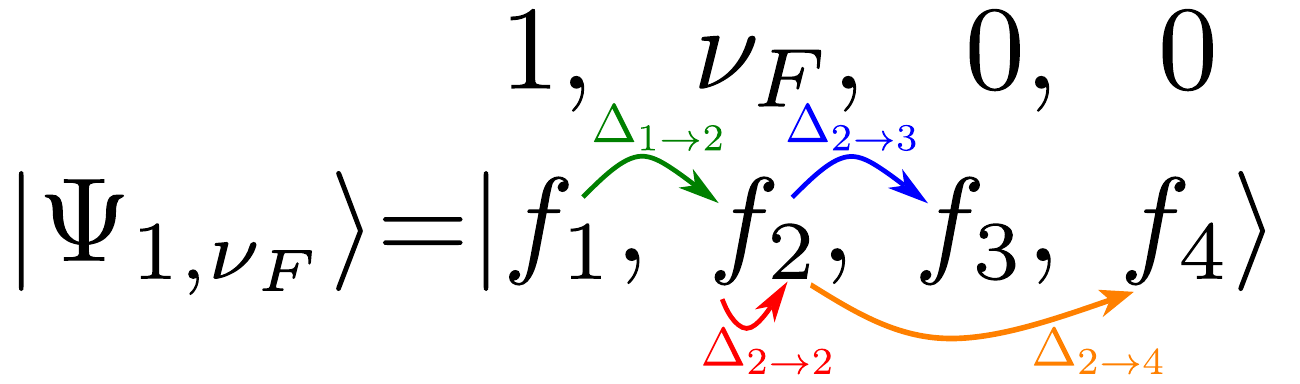}
    \caption{The excitations relevant to the transport gap. We assume that spinor/flavor $|f_1\rangle$ is fully filled, $|f_2\rangle$ has filling $\nu_F$, and  $|f_3\rangle$ and $|f_4\rangle$ are empty. The four possible excitations involving $|f_2\rangle$ are shown, labeled as $\Delta_{i\rightarrow j}$ in a self-evident notation. Of these, $\Delta_{2\rightarrow 2}$ corresponds to the unique flavor-preserving gap, while the others change the flavor.}
    \label{fig_all_gaps}
\end{figure}

A key point here is that transport measurements are sensitive to the charge excitations and depend on both the ground and excited states. By contrast, local STM measurements directly probe lattice symmetry-breaking due to valley-ordering. In typical semiconductor systems, the spin degeneracy is strongly lifted at high $B$-fields and only the spin/flavor-conserving charge excitations are relevant at low temperatures. On the other hand, in graphene, the (nearly) degenerate set of four LLs leads to several possible excitations [see Fig.~\ref{fig_all_gaps}] which could be close in energy. If a flavor-changing excitation is the lowest in energy, it will dominate the transport gap. 
In this Letter, we propose a theoretical explanation for the puzzling observations of Ref.~\cite{GoldhaberGordon2015} on the contrasting dependence of the transport gaps in the ZLLs and 1LLs. Our analysis is consistent with all known experimental results and constrains the values of the microscopic anisotropic couplings. Using these constraints, we also make predictions for the nature of the ground state and the Zeeman dependence of the transport gap for fractions that have not been investigated in detail previously. 

The root cause of the difference between the ZLLs and the 1LLs is that the projection of a given set of microscopic interactions to the two manifolds results in different effective Hamiltonians because (i) some microscopic couplings disappear on projection to the ZLLs due to valley-sublattice locking, and (ii) the effective couplings in the 1LL manifold acquire a nonzero range due to the form of the wavefunctions even if the microscopic interactions are USR. Thus, one expects to see different phases at $\nu{=}{-}1/3~(\text{or }{-}2/3)$ and $\nu{=}4{-}1/3~(\text{or }4{-}2/3)$ for the same microscopic couplings~\cite{Sodemann_MacDonald_2014, Jincheng2024, an2024fractional}. By carefully analyzing \textit{all} possible low-lying charge excitations in the various phases of the $\bn{=}0,1$ manifolds, we find a range of microscopic couplings that explains the anomalous $E_Z$ dependence of the 1LL transport gap~\cite{GoldhaberGordon2015}.

{\bf \em Hamiltonian, Energy Functional, Gaps---} We start with a USR anisotropic interacting Hamiltonian 
\begin{equation}
\hat H^\text{an}=\dfrac{1}{2}\sum_{a,b}g_{ab}\int d^2\br:\Big[\hat\psi^\dagger(\br)\Big(\tau^a\otimes\eta^b\Big)\hat\psi(\br)\Big]^2:,
\end{equation}
where $\tau~(\eta)$ are Pauli matrices in valley (sublattice) space with $a,b{=}0,x,y,z$. Constrained by the honeycomb lattice symmetries of graphene, the couplings $g_{ab}$ satisfy  relations~\cite{RG1_Graphene_Aleiner_2007, RG2_Graphene_Aleiner_2008, kharitonov2012:nu0} $g_{ax}{=}g_{ay}{\equiv}g_{a \perp}$ and $g_{xb}{=}g_{yb}{\equiv}g_{\perp b}$. It is very convenient to use the Haldane pseudopotentials~\cite{Haldane_Pseudopot1983} to characterize the anisotropies. The pseudopotential $u^{(m)}_{a,\bn}$ is the interaction amplitude between electrons of relative angular momentum $m$, in the anisotropy channel $a$, projected to the $\bn$-th LL manifold. A USR interaction has only the $m{=}0$ pseudopotentials nonzero. 
After projection to the $\bn$-th LL manifold, the Hamiltonian becomes~\cite{an2024fractional}  
\bean\hspace*{-0.5cm}\label{eq_projected_Hamiltonian}
&&\hat H^\text{an}_{\bn}=\dfrac{1}{2}\sum_m\sum_{a=x,y,z} u^{(m)}_{a,\bn}\bU^{(m)}_{m_1m_2m_3m_4}:\hat\tau^a_{m_1m_4}\hat\tau^a_{m_2m_3}:,
\eean
where $\hat\tau^a_{m_1m_4}{=}(\tau^a{\otimes} \sigma^0_{\text{spin}} )_{\alpha\beta}\hcd_{m_1,\alpha}\hc_{m_4,\beta}$ with ${\alpha},{\beta} {=} {K{\uparrow}},\ K{\downarrow},\ K^\prime{\uparrow},\ K^\prime{\downarrow}$, and the interactions have been decomposed into Haldane pseudopotentials $u^{(m)}_{x,\bn}{=}u^{(m)}_{y,\bn}{\equiv} u^{(m)}_{\perp,\bn}$, $u^{(m)}_{z,\bn}$ with corresponding interaction matrix element $\bU^{(m)}_{m_1m_2m_3m_4}$.
\bean\label{eq_haldane_pseudopotentials}\hspace*{-0.7cm}
\quad u^{(0)}_{a,\bn = 0}= \frac{g_{a 0} + g_{a z}}{4 \pi \ell^2},\ 
\begin{cases}\label{n1_haldane}
   u^{(0)}_{a,\bn = 1} = \displaystyle\frac{5 g_{a 0} + g_{a z} + 4 g_{a \perp}}{32 \pi \ell^2}, \\[10pt]
    u^{(1)}_{a,\bn = 1} =\displaystyle \frac{g_{a0} - g_{a z} - 2 g_{a \perp}}{16 \pi \ell^2}, \\[10pt] 
    u^{(2)}_{a,\bn = 1} = \displaystyle\frac{g_{a 0} + g_{a z}}{32 \pi \ell^2},
\end{cases}
\eean
In total, there are six independent anisotropic couplings.\\

Following earlier work~\cite{Sodemann_MacDonald_2014} and inputs from exact diagonalization, we derive a variational energy functional [see Refs.~\cite{Jincheng2024, an2024fractional} and supplemental material (SM)~\cite{SM}] for FQH states $|\Psi\rgl{=}|1f_1,\nu_Ff_2\rgl$ occupying spinors $|f_1\rgl\ \&\ |f_2\rgl$ at filling $(1,\nu_F)$ with $\nu_F{=}1/3,~2/3$. With projectors to the filled spinors, $P_I{=}|f_1\rgl\lgl f_1|$ and $ P_F{=}|f_{2}\rgl\lgl f_{2}|$, the energy functional is schematically written as  
$\lgl\Psi|\hat H^\text{an}_\bn|\Psi\rgl{\sim} E_{II}(P_I, P_I){+}E_{IF}(P_I, P_F){+}E_{FF}(P_F, P_F)$, where $II$ stands for the self-interaction of the integer-filled flavors, $IF$ for the interaction between the integer and fractional flavors, and $FF$ for the self-interaction of the fractionally filled flavor. The explicit expressions are provided in the SM~\cite{SM}. The energy functional, including the Zeeman term ${-}E_Z\sigma^z_\text{spin}$ with $E_Z{=}(1/2)g_{B}\mu_BB_\text{tot}$(the $g$-factor $g_{B}{=}2$ for graphene), can be variationally minimized using the spinor ansatz proposed in Refs.~\cite{an2024fractional, Jincheng2024}.

Once the ground state has been obtained, we compute the transport gaps associated with the fractionally filled spinor $|f_2\rgl$. As shown in Fig.~\ref{fig_all_gaps}, these include $\Delta_{2{\rightarrow}i}$ with $i{=}2,3,4$, arising from a quasi-hole in $|f_2\rgl$ and a quasi-particle in $|f_i\rgl$ and $\Delta_{1{\rightarrow}2}$ corresponding to a quasi-hole in $|f_1\rgl$ a quasi-particle in $|f_2\rgl$. The three gaps $\Delta_{i{\rightarrow} j{\neq} i}$ arise from flavor-flip processes, while $\Delta_{2{\rightarrow}2}$ is a flavor-conserving gap. The gaps are composed of three contributions, $\Delta_{i{\rightarrow} j}{\equiv}\Delta_C{+}\Delta_{1b}{+}\Delta_F$, where $\Delta_C$ is the Coulomb contribution that is evaluated by exact diagonalization~\cite{Morf02, Balram20b, SM}, $\Delta_{1b}{=}\lgl f_j|H_{1b}|f_j\rgl{-}\lgl f_i| H_{1b}|f_i\rgl$ is due to one-body terms $H_{1b}$ including the Zeeman energy and the anisotropic interaction of the excitation with the integer-filled spinor(s)~\cite{Sodemann_MacDonald_2014}, 
\bean
&&H_{1b}=\sum_{a}\Big[u_{a, H}\rm{Tr}\big(P_I\tau^a\big)\tau^a-u_{a,F}\tau^aP_I\tau^a\Big]-E_Z\sigma^z,\nn
&&u_{a,H}=2\sum_mu_a^{(m)},\quad u_{a,H}=2\sum_m(-1)^mu_a^{(m)},
\eean
and $\Delta_F$ accounts for contributions from the fractional sector via the anisotropic interaction. The evaluation of these gaps is described in the SM~\cite{SM}.

\begin{figure}[h]
    \centering \includegraphics[width=0.48\textwidth,height=0.45\textwidth]{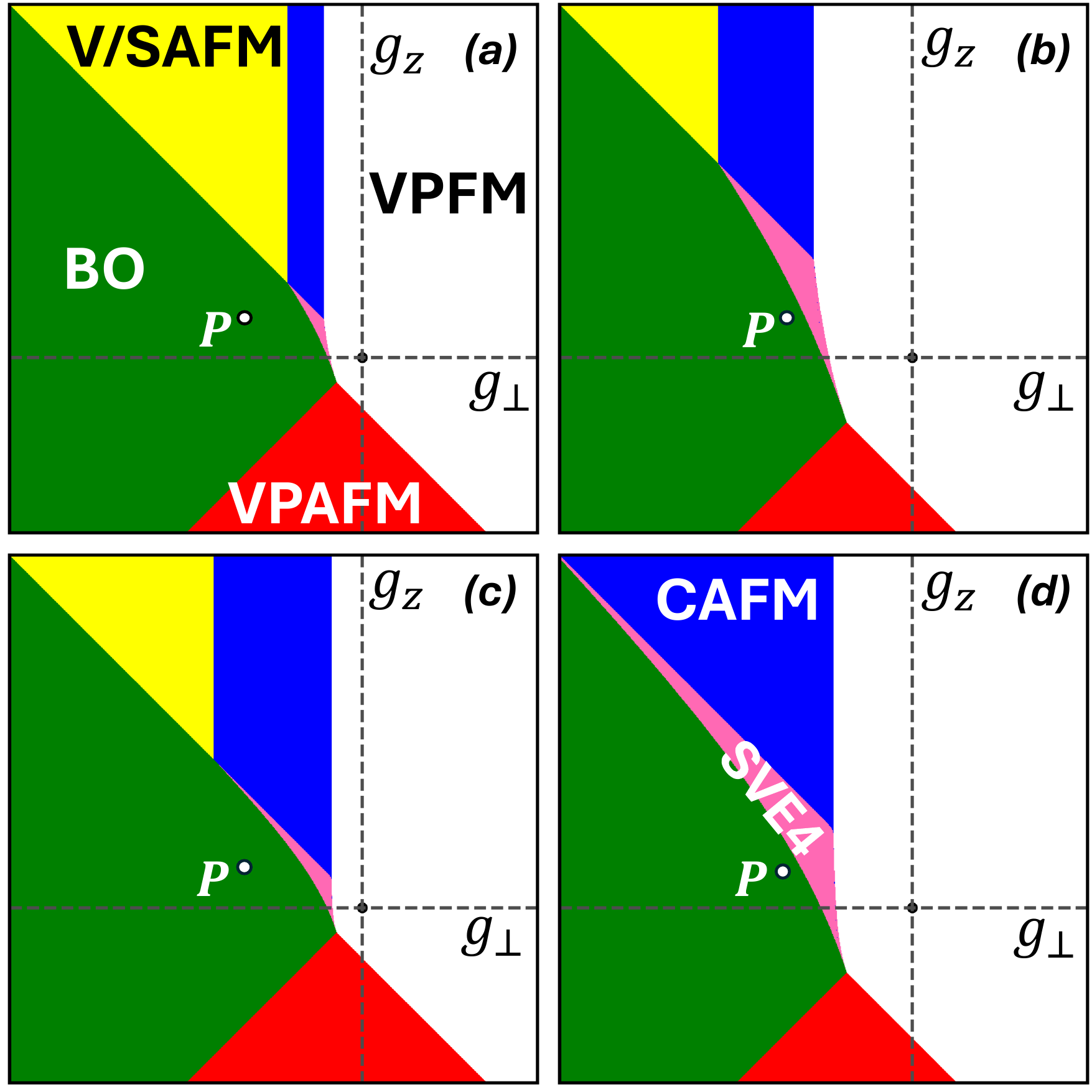}
    \caption{Phase diagrams for the $(1,\nu_F)$ state in the ZLL with $ g_{\perp}{=}g_{\perp 0}{+}g_{\perp z}{\in} [{-}1600,800]\text{meV}{\cdot} \text{nm}^{2}$, $g_{z}{=}g_{z0}{+}g_{zz}{\in}[{-}800,1600]\text{meV}{\cdot}\text{nm}^{2}$. (a) $\nu_F{=}1/3$, $B_\text{tot}{=}B_\perp{=}17\text{T}$, (b) $\nu_F{=}1/3,$ $B_\text{tot}{=}44\text{T}$, $B_\perp{=}17\text{T}$, (c) $\nu_F{=}2/3,\ B_\text{tot}{=}B_\perp{=}17\text{T}$, (d) $\nu_F{=}2/3,$ $B_\text{tot}{=}44\text{T}$, $B_\perp{=}17\text{T}$. At point $P$: $(g_\perp,\ g_z){=}({-}600,\ 200)\text{meV}{\cdot} \text{nm}^{2}$, and the ground states of both fillings remain in the BO phase as $B_\text{tot}$ increases from 17 to 44 T.
    }
    \label{fig_n0_phase_diagrams}
\end{figure}

{\bf \em Phase Diagrams, Transport Gaps---} 
In ZLL, as can be seen in Eq.~\eqref{eq_haldane_pseudopotentials}, the anisotropic interactions are quantified by two ``effective" couplings $g_\perp{\equiv} g_{\perp 0}{+}g_{\perp z}$ and $g_z{\equiv} g_{z0}{+}g_{zz}$. In Fig.~\ref{fig_n0_phase_diagrams}, we present the phase diagrams for FQH states at filling $(1,\nu_F)$ with $\nu_F{=}1/3,~2/3$ in the ZLL at two different values of $B_\text{tot}$. We see the spin-valley entangled (SVE) phase SVE4, with the occupied spinors $ |f_1\rgl{=}\cos({\alpha_1}/{2})|\hat e_x\rgl{\otimes}|{\uparrow}\rangle{-}\sin({\alpha_1}/{2})|{-}\hat e_x\rgl{\otimes}|{\downarrow}\rangle$,\  $|f_2\rgl{=}\cos({\alpha_2}/{2})|\hat e_x\rgl{\otimes}|{\downarrow}\rangle{-}\sin({\alpha_2}/{2})|{-}\hat e_x\rgl{\otimes}|{\uparrow}\rangle$. These spinors cannot be written as a direct product of 2-dimensional spin and valley spinors, hence the label.\\

\begin{figure}[h]
    \centering \includegraphics[width=0.48\textwidth,height=0.47\textwidth]{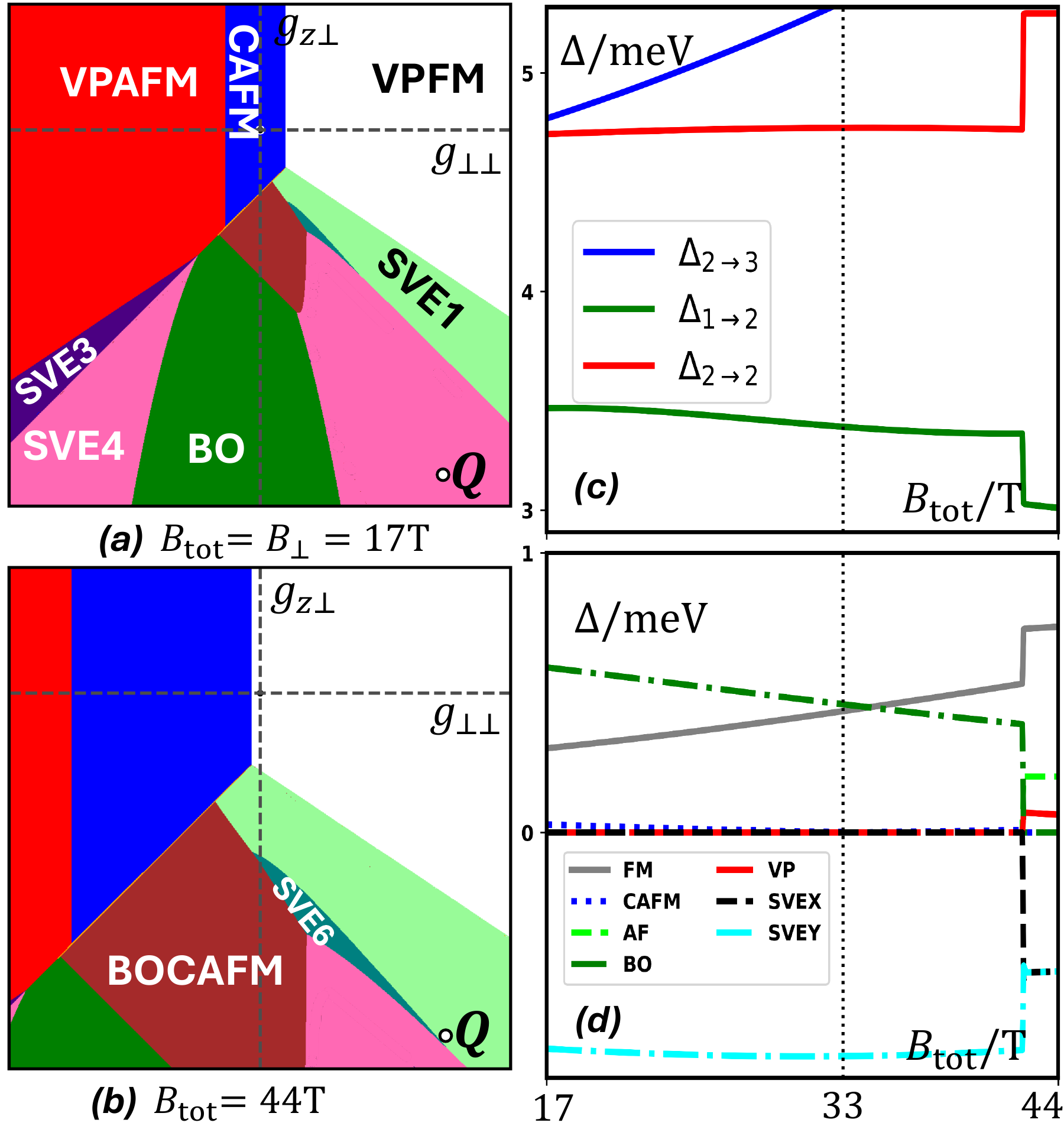}
    \caption{(a-b) $(1,\frac{2}{3})$ phase diagrams in 1LL with $ g_{\perp\perp}{\in}[{-}1600,1600]\text{meV}{\cdot} \text{nm}^{2}$, $ g_{z\perp}{\in} [{-}2400,800]\text{meV}{\cdot} \text{nm}^{2}$ at two different $B_\text{tot}$. (c) Transport gaps $\Delta$'s and (d) order parameters with increasing $B_\text{tot}$ from 17T to 44T at point $Q$: $(g_{\perp\perp},\ g_{z\perp}){=}(1180,\ {-}2200)\text{meV}{\cdot} \text{nm}^{2}$. A first-order phase transition from phase SVE4 to SVE1 occurs as $E_Z$ increases. }
    \label{fig_13200_gaps}
\end{figure}

Since only the USR Haldane pseudopotential $u^{(0)}_{a,\bn{=}0}$ is relevant in the ZLL, and ground and excited states are described by hard-core wavefunctions~\cite{Wu93}, $\Delta_F{=}0$ for the all the gaps $\Delta_{i{\rightarrow} j}$. At  point $P$: $(g_\perp,\ g_z){=}({-}600,\ 200)\text{meV}{\cdot} \text{nm}^{2}$ highlighted in Fig.~\ref{fig_n0_phase_diagrams}(a-d), the ground state is in the \BOcircle\ BO phase described by the spinors $|f_1\rgl{=}|e_x\rgl{\otimes}|{\uparrow}\rgl,\ |f_2\rgl{=}|e_x\rgl{\otimes}|{\downarrow}\rgl$ (with $ {\pm} e_x{=}(K{\pm} K^\prime)/\sqrt{2}$) for both $(1,1/3)$ and $(1,2/3)$ across the entire range of $E_Z$. The various gaps of Fig.~\ref{fig_all_gaps} for $(1,1/3)$ in the BO phase can be computed analytically, and are
\bean\label{eq_n0_bo_gap_4third}
&&\Delta_{2\rightarrow2}=0.10\dfrac{e^2}{\kappa\ell},\ \Delta_{2\rightarrow3}=0.075\dfrac{e^2}{\kappa\ell}-\dfrac{-3g_\perp-g_z}{2\pi\ell^2}-2E_Z,\nn  &&\Delta_{2\rightarrow4}=0.075\dfrac{e^2}{\kappa\ell}-\dfrac{g_\perp}{\pi\ell^2},\ \Delta_{1\rightarrow2}=0.067\dfrac{e^2}{\kappa\ell}+2E_Z.
\eean 
The lowest gap is $\Delta_{2{\rightarrow}2}{=}0.10{e^2}/{\kappa\ell}$ with $\kappa{\approx}6\epsilon_{0}$ being the dielectric constant of graphene~\cite{Hunt2016DirectMO}. Note that $\Delta_{2{\rightarrow}2}$ is independent of $E_Z$ and is thus constant when $B_{\parallel}$ is varied. The ground state at the filling $(1,2/3)$ also remains in the BO phase for all the $E_Z$ we considered, and also has the lowest gap $\Delta_{2{\rightarrow}2}$.
\begin{figure}[h]
    \centering \includegraphics[width=0.48\textwidth,height=0.47\textwidth]{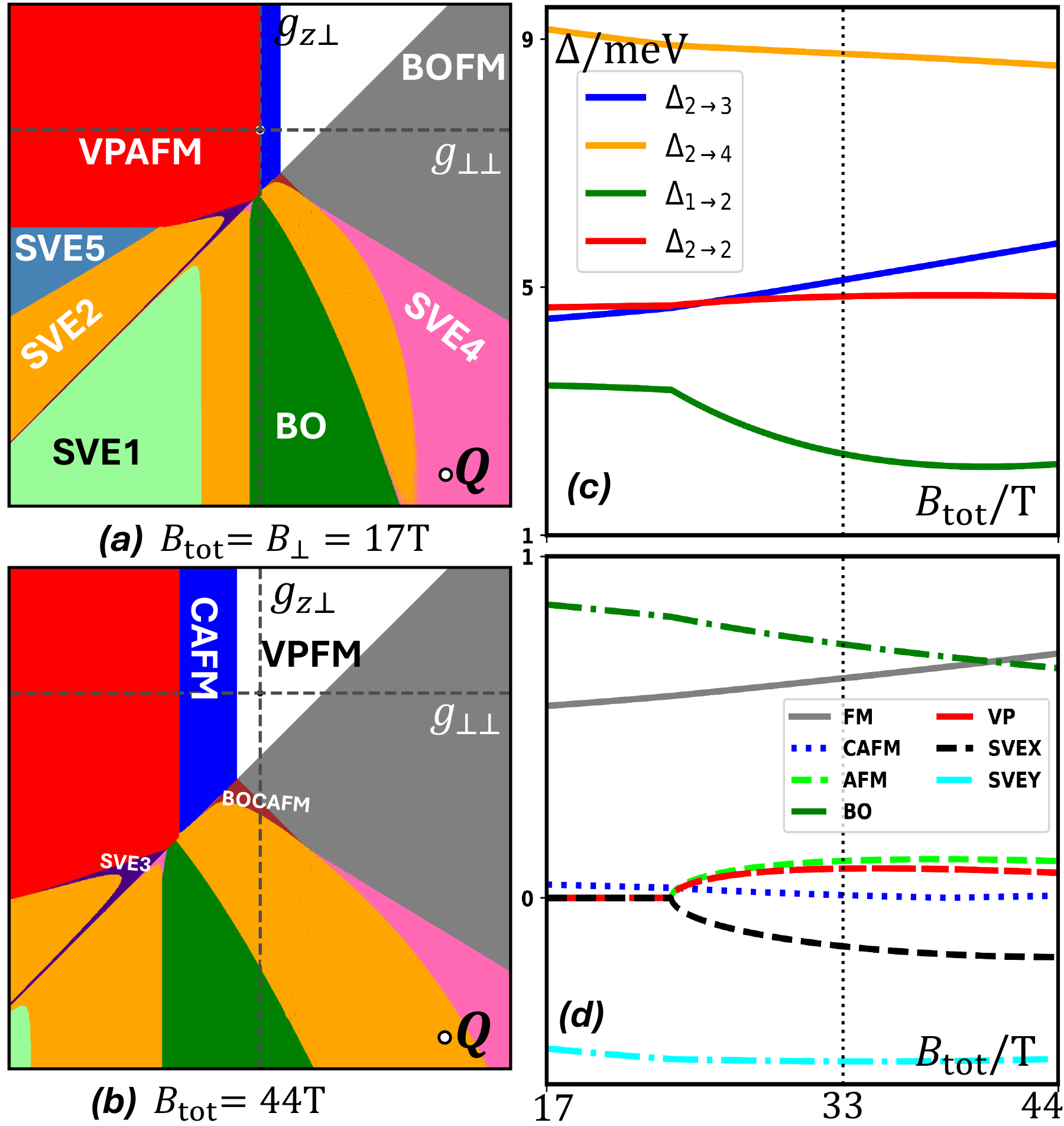}
    \caption{(a-b) $(1,1/3)$ phase diagrams in 1LL with $ g_{\perp\perp}{\in} [{-}1600,1600]\text{meV}{\cdot} \text{nm}^{2}$, $ g_{z\perp}{\in} [{-}2400,800]\text{meV}{\cdot} \text{nm}^{2}$ at two different $B_\text{tot}$. (c) Transport gaps $\Delta$'s and (d) Order parameters with increasing $B_\text{tot}$ from 17T to 44T at point $Q$: $(g_{\perp\perp},\ g_{z\perp}){=}(1180,\ {-}2200)\text{meV}{\cdot} \text{nm}^{2}$. A second-order phase transition from phase SVE4 to SVE2 happened as the Zeeman energy increased.}
    \label{fig_13100_gaps}
\end{figure}
As discussed in Refs.~\cite{kharitonov2012:nu0, Wei_Xu_Sodemann_Huang_LLM_SU4_breaking_MLG_2024}, microscopic estimates indicate that $g_{\perp 0} {=} g_{z0} {=} 0$. We use these values, and also fix $(g_{\perp z},\ g_{zz})$ at the point labeled $P$ in Fig.~\ref{fig_n0_phase_diagrams}. The remaining anisotropic couplings are $g_{\perp\perp} $ and $g_{z\perp}$. These couplings project to zero in the ZLLs due to valley-sublattice locking, but play an important role in the 1LLs. In Figs.~\ref{fig_13100_gaps}(a-b) and~\ref{fig_13200_gaps}(a-b), we present the phase diagrams for the FQH states of fillings $(1,1/3)$ and $(1,2/3)$ in the 1LLs in the $(g_{\perp\perp},\ g_{z\perp})$ parameter space. Owing to the three leading Haldane pseudopotentials being nonzero in the 1LL [see Eq.~\eqref{eq_haldane_pseudopotentials}], additional phases emerge that do not occur in the ZLLs. The details of these phases are presented in the SM~\cite{SM}.\\

After a thorough exploration of the parameter space, we find that in the neighborhood of $Q$: $(g_{\perp\perp},\ g_{z\perp}) {=} (1180,\ {-}2200),\text{meV}{\cdot} \text{nm}^2$, the lowest transport gaps decrease vs. $E_Z$, as shown in Figs.~\ref{fig_13100_gaps}(c) and ~\ref{fig_13200_gaps}(c). At the lowest $E_Z$ ($B_{\parallel}=0$), the ground states of both fillings remain in the \twoCSsveFOUR\ SVE4 (spin-valley entangled) phase with the spinors $ |f_1\rgl{=}\cos({\alpha_1}/{2})|\hat e_x\rgl{\otimes}|{\uparrow}\rangle{-}\sin({\alpha_1}/{2})|{-}\hat e_x\rgl{\otimes}|{\downarrow}\rangle$,\  $|f_2\rgl{=}\cos({\alpha_2}/{2})|\hat e_x\rgl{\otimes}|{\downarrow}\rangle{-}\sin({\alpha_2}/{2})|{-}\hat e_x\rgl{\otimes}|{\uparrow}\rangle$. For the filling $(1,1/3)$, as shown in Fig.~\ref{fig_13100_gaps}(d), increasing $E_Z$  induces a second-order phase transition into the \twoCSsveTWO\ SVE2 phase, in which the spinors are very general superpositions of valley-spin basis $|\btau\rgl{\otimes}|\bs\rgl$ and all order parameters (OPs) are nonzero. In Fig.~\ref{fig_13200_gaps}(d), we see that $(1,2/3)$ will enter the \twoCSsveONE\ SVE1 phases with spinors $|f_1\rgl{=}\cos{({\alpha}/{2})}|K^\prime \rgl{\otimes}|{\uparrow}\rangle{-} \sin({\alpha}/{2})|K\rgl{\otimes}|{\downarrow}\rangle,\ |f_2\rgl{=}|K\rgl{\otimes}|{\uparrow}\rangle$ via a first-order transition.\\

Finally, at filling $\nu{=}2{+}2/3$, only one of the 4 spinors in the 1LL manifold is 2/3-filled. The FQH state has only a single flavor and responds only to $u^{(1)}_{a,\bn{=}1}$. The resulting phase diagram is the same as that of the $\nu{=}3$ state~\cite{Lian_Goerbig_2017, Atteia_Goerbig_2021}. Using the couplings we specified above, we obtain the optimal spinor $|f\rgl{=}|e_x\rgl{\otimes}|{\uparrow}\rgl$ for $\nu{=}2{+}2/3$. This shows that the transport gap is independent of $E_Z$, consistent with experiment~\cite{GoldhaberGordon2015}. \\

{\bf \em Conclusions and Outlook---} It has been known for a decade that the FQH phases that occur in the 1LLs of graphene have a very different $E_Z$ dependence of the transport gap~\cite{GoldhaberGordon2015} from those that occur in the ZLLs.  Our explanation of this phenomenon relies on quantum Hall ferromagnetism, which is driven by a combination of the long-range Coulomb, short-range anisotropic interactions, and one-body couplings. Due to form factor differences between the ZLLs and the 1LLs, the effective Hamiltonian is quite different in the two LL manifolds, leading to very different spin/valley symmetry-breaking  ground states. To compare transport gaps, in addition to the ground state, one also needs to determine which excitation is the lowest. Using well-known variational wavefunctions and exact diagonalization~\cite{Laughlin_1983, Jain89, Morf02, Sodemann_MacDonald_2014, Balram20b, Jincheng2024, an2024fractional}, we have carried out this program and found a set of microscopic couplings that are consistent with previous measurements on integer~\cite{young2014:nu0} and fractional states~\cite{GoldhaberGordon2015} in the $\bn{=}0,1$ LL manifolds. Notably, the 1LL states exhibit an exotic spin-valley entangled order, meaning the occupied spinors cannot be written as direct products of spin and valley spinors.  \\

Using the constrained couplings, we can make predictions: (i) In the 1LL, the state $(1,2/3)$ should display bond order at low $E_Z$ but none at high $E_Z$, due to a ground state phase transition from SVE4 to SVE1. (ii) STM in a tunneling configuration can pick up the effective LLs of composite fermions by injecting an electron~\cite{STM_Yazdani2021visualizing,STM_Coissard_2022, Farahi23, Pu22, Pu23a, Gattu23}. All states with an extra electronic charge, whether flavor-conserving or flavor-flipping, should contribute to this spectrum, which can be computed in our approach. Such an experiment would produce a much more detailed fingerprint of each phase. (iii)  In principle, bulk thermal transport measurements should allow us to deduce whether a given phase has Goldstone modes~\cite{Anindya_Das_2024_Heat_transport, Parmentier_2024_Heat_Flow}. We find that the ZLL phases at the fractions measured are all bond-ordered and should not have gapless modes, whereas the 1LL phases are all spin-valley entangled and have gapless Goldstone modes.\\

Apart from the specific application to FQH states in monolayer graphene, our approach is potentially applicable to other problems where multiple internal quantum degrees of freedom lead to QHFM. The ideas presented here may be fruitful in analyzing the transport gaps in the FQH regime of multilayer graphene systems~\cite{Zibrov16, Li17, HUang21, Assouline23, Kumar24, Chen24, Chanda25}, as well as the anomalous FQH regime in moir\'e and other systems~\cite{FQAH_MoTe2_Xu_2023a, FQAH_MoTe2_Xu_2023b, FQAH_MoTe2_Mak_Shan_2023, FQAHE_MoTe2_Li_2023, FQAH_Pentalayer_Graphene_Ju_2024}.

{\bf \em Acknowledgements.---}
J.A. and G.M. are partially supported by the U.S. Department of Energy, Office of Science, Office of Basic Energy Sciences under Award Number DE-SC0024346. J.A. is also grateful to the University of Kentucky Center for Computational Sciences and Information Technology Services Research Computing for allowing the use of the Morgan Compute Cluster. U.K. acknowledges support from the Department of Space (DOS), Government of India. We acknowledge the Science and Engineering Research Board (SERB) of the Department of Science and Technology (DST) for financial support through the Mathematical Research Impact Centric Support (MATRICS) Grant No. MTR/2023/000002. Some of the numerical calculations reported in this work were carried out on the Nandadevi and Kamet supercomputers, which are maintained and supported by the Institute of Mathematical Science’s High-Performance Computing Center. Some numerical calculations were performed using the DiagHam libraries~\cite{diagham} for which we are grateful to the authors.

\bibliography{hall}

\begin{thebibliography}{67}%
\makeatletter
\providecommand \@ifxundefined [1]{%
 \@ifx{#1\undefined}
}%
\providecommand \@ifnum [1]{%
 \ifnum #1\expandafter \@firstoftwo
 \else \expandafter \@secondoftwo
 \fi
}%
\providecommand \@ifx [1]{%
 \ifx #1\expandafter \@firstoftwo
 \else \expandafter \@secondoftwo
 \fi
}%
\providecommand \natexlab [1]{#1}%
\providecommand \enquote  [1]{``#1''}%
\providecommand \bibnamefont  [1]{#1}%
\providecommand \bibfnamefont [1]{#1}%
\providecommand \citenamefont [1]{#1}%
\providecommand \href@noop [0]{\@secondoftwo}%
\providecommand \href [0]{\begingroup \@sanitize@url \@href}%
\providecommand \@href[1]{\@@startlink{#1}\@@href}%
\providecommand \@@href[1]{\endgroup#1\@@endlink}%
\providecommand \@sanitize@url [0]{\catcode `\\12\catcode `\$12\catcode `\&12\catcode `\#12\catcode `\^12\catcode `\_12\catcode `\%12\relax}%
\providecommand \@@startlink[1]{}%
\providecommand \@@endlink[0]{}%
\providecommand \url  [0]{\begingroup\@sanitize@url \@url }%
\providecommand \@url [1]{\endgroup\@href {#1}{\urlprefix }}%
\providecommand \urlprefix  [0]{URL }%
\providecommand \Eprint [0]{\href }%
\providecommand \doibase [0]{https://doi.org/}%
\providecommand \selectlanguage [0]{\@gobble}%
\providecommand \bibinfo  [0]{\@secondoftwo}%
\providecommand \bibfield  [0]{\@secondoftwo}%
\providecommand \translation [1]{[#1]}%
\providecommand \BibitemOpen [0]{}%
\providecommand \bibitemStop [0]{}%
\providecommand \bibitemNoStop [0]{.\EOS\space}%
\providecommand \EOS [0]{\spacefactor3000\relax}%
\providecommand \BibitemShut  [1]{\csname bibitem#1\endcsname}%
\let\auto@bib@innerbib\@empty
\bibitem [{\citenamefont {Tsui}\ \emph {et~al.}(1982)\citenamefont {Tsui}, \citenamefont {Stormer},\ and\ \citenamefont {Gossard}}]{FQHE_Discovery_1982}%
  \BibitemOpen
  \bibfield  {author} {\bibinfo {author} {\bibfnamefont {D.~C.}\ \bibnamefont {Tsui}}, \bibinfo {author} {\bibfnamefont {H.~L.}\ \bibnamefont {Stormer}},\ and\ \bibinfo {author} {\bibfnamefont {A.~C.}\ \bibnamefont {Gossard}},\ }\bibfield  {title} {\bibinfo {title} {Two-dimensional magnetotransport in the extreme quantum limit},\ }\href {https://doi.org/10.1103/PhysRevLett.48.1559} {\bibfield  {journal} {\bibinfo  {journal} {Phys. Rev. Lett.}\ }\textbf {\bibinfo {volume} {48}},\ \bibinfo {pages} {1559} (\bibinfo {year} {1982})}\BibitemShut {NoStop}%
\bibitem [{\citenamefont {Laughlin}(1983)}]{Laughlin_1983}%
  \BibitemOpen
  \bibfield  {author} {\bibinfo {author} {\bibfnamefont {R.~B.}\ \bibnamefont {Laughlin}},\ }\bibfield  {title} {\bibinfo {title} {Anomalous quantum {Hall} effect: An incompressible quantum fluid with fractionally charged excitations},\ }\href {https://doi.org/10.1103/PhysRevLett.50.1395} {\bibfield  {journal} {\bibinfo  {journal} {Phys. Rev. Lett.}\ }\textbf {\bibinfo {volume} {50}},\ \bibinfo {pages} {1395} (\bibinfo {year} {1983})}\BibitemShut {NoStop}%
\bibitem [{\citenamefont {Arovas}\ \emph {et~al.}(1984)\citenamefont {Arovas}, \citenamefont {Schrieffer},\ and\ \citenamefont {Wilczek}}]{Arovas84}%
  \BibitemOpen
  \bibfield  {author} {\bibinfo {author} {\bibfnamefont {D.}~\bibnamefont {Arovas}}, \bibinfo {author} {\bibfnamefont {J.~R.}\ \bibnamefont {Schrieffer}},\ and\ \bibinfo {author} {\bibfnamefont {F.}~\bibnamefont {Wilczek}},\ }\bibfield  {title} {\bibinfo {title} {Fractional statistics and the quantum {Hall} effect},\ }\href {https://doi.org/10.1103/PhysRevLett.53.722} {\bibfield  {journal} {\bibinfo  {journal} {Phys. Rev. Lett.}\ }\textbf {\bibinfo {volume} {53}},\ \bibinfo {pages} {722} (\bibinfo {year} {1984})}\BibitemShut {NoStop}%
\bibitem [{\citenamefont {Fertig}(1989)}]{QHFM_Fertig_1989}%
  \BibitemOpen
  \bibfield  {author} {\bibinfo {author} {\bibfnamefont {H.~A.}\ \bibnamefont {Fertig}},\ }\bibfield  {title} {\bibinfo {title} {Energy spectrum of a layered system in a strong magnetic field},\ }\href {https://doi.org/10.1103/PhysRevB.40.1087} {\bibfield  {journal} {\bibinfo  {journal} {Phys. Rev. B}\ }\textbf {\bibinfo {volume} {40}},\ \bibinfo {pages} {1087} (\bibinfo {year} {1989})}\BibitemShut {NoStop}%
\bibitem [{\citenamefont {Sondhi}\ \emph {et~al.}(1993)\citenamefont {Sondhi}, \citenamefont {Karlhede}, \citenamefont {Kivelson},\ and\ \citenamefont {Rezayi}}]{Shivaji_Skyrmion}%
  \BibitemOpen
  \bibfield  {author} {\bibinfo {author} {\bibfnamefont {S.~L.}\ \bibnamefont {Sondhi}}, \bibinfo {author} {\bibfnamefont {A.}~\bibnamefont {Karlhede}}, \bibinfo {author} {\bibfnamefont {S.~A.}\ \bibnamefont {Kivelson}},\ and\ \bibinfo {author} {\bibfnamefont {E.~H.}\ \bibnamefont {Rezayi}},\ }\bibfield  {title} {\bibinfo {title} {Skyrmions and the crossover from the integer to fractional quantum {Hall} effect at small {Zeeman} energies},\ }\href {https://doi.org/10.1103/PhysRevB.47.16419} {\bibfield  {journal} {\bibinfo  {journal} {Phys. Rev. B}\ }\textbf {\bibinfo {volume} {47}},\ \bibinfo {pages} {16419} (\bibinfo {year} {1993})}\BibitemShut {NoStop}%
\bibitem [{\citenamefont {Yang}\ \emph {et~al.}(1994)\citenamefont {Yang}, \citenamefont {Moon}, \citenamefont {Zheng}, \citenamefont {MacDonald}, \citenamefont {Girvin}, \citenamefont {Yoshioka},\ and\ \citenamefont {Zhang}}]{QHFM_Yang_etal_1994}%
  \BibitemOpen
  \bibfield  {author} {\bibinfo {author} {\bibfnamefont {K.}~\bibnamefont {Yang}}, \bibinfo {author} {\bibfnamefont {K.}~\bibnamefont {Moon}}, \bibinfo {author} {\bibfnamefont {L.}~\bibnamefont {Zheng}}, \bibinfo {author} {\bibfnamefont {A.~H.}\ \bibnamefont {MacDonald}}, \bibinfo {author} {\bibfnamefont {S.~M.}\ \bibnamefont {Girvin}}, \bibinfo {author} {\bibfnamefont {D.}~\bibnamefont {Yoshioka}},\ and\ \bibinfo {author} {\bibfnamefont {S.-C.}\ \bibnamefont {Zhang}},\ }\bibfield  {title} {\bibinfo {title} {Quantum ferromagnetism and phase transitions in double-layer quantum {Hall} systems},\ }\href {https://doi.org/10.1103/PhysRevLett.72.732} {\bibfield  {journal} {\bibinfo  {journal} {Phys. Rev. Lett.}\ }\textbf {\bibinfo {volume} {72}},\ \bibinfo {pages} {732} (\bibinfo {year} {1994})}\BibitemShut {NoStop}%
\bibitem [{\citenamefont {Moon}\ \emph {et~al.}(1995)\citenamefont {Moon}, \citenamefont {Mori}, \citenamefont {Yang}, \citenamefont {Girvin}, \citenamefont {MacDonald}, \citenamefont {Zheng}, \citenamefont {Yoshioka},\ and\ \citenamefont {Zhang}}]{QHFM_Moon_etal_1995}%
  \BibitemOpen
  \bibfield  {author} {\bibinfo {author} {\bibfnamefont {K.}~\bibnamefont {Moon}}, \bibinfo {author} {\bibfnamefont {H.}~\bibnamefont {Mori}}, \bibinfo {author} {\bibfnamefont {K.}~\bibnamefont {Yang}}, \bibinfo {author} {\bibfnamefont {S.~M.}\ \bibnamefont {Girvin}}, \bibinfo {author} {\bibfnamefont {A.~H.}\ \bibnamefont {MacDonald}}, \bibinfo {author} {\bibfnamefont {L.}~\bibnamefont {Zheng}}, \bibinfo {author} {\bibfnamefont {D.}~\bibnamefont {Yoshioka}},\ and\ \bibinfo {author} {\bibfnamefont {S.-C.}\ \bibnamefont {Zhang}},\ }\bibfield  {title} {\bibinfo {title} {Spontaneous interlayer coherence in double-layer quantum {Hall} systems: Charged vortices and {Kosterlitz}-{Thouless} phase transitions},\ }\href {https://doi.org/10.1103/PhysRevB.51.5138} {\bibfield  {journal} {\bibinfo  {journal} {Phys. Rev. B}\ }\textbf {\bibinfo {volume} {51}},\ \bibinfo {pages} {5138} (\bibinfo {year} {1995})}\BibitemShut {NoStop}%
\bibitem [{\citenamefont {Novoselov}\ \emph {et~al.}(2004)\citenamefont {Novoselov}, \citenamefont {Geim}, \citenamefont {Morozov}, \citenamefont {Jiang}, \citenamefont {Zhang}, \citenamefont {Dubonos}, \citenamefont {Grigorieva},\ and\ \citenamefont {Firsov}}]{Novoselov_etal_2004}%
  \BibitemOpen
  \bibfield  {author} {\bibinfo {author} {\bibfnamefont {K.~S.}\ \bibnamefont {Novoselov}}, \bibinfo {author} {\bibfnamefont {A.~K.}\ \bibnamefont {Geim}}, \bibinfo {author} {\bibfnamefont {S.~V.}\ \bibnamefont {Morozov}}, \bibinfo {author} {\bibfnamefont {D.}~\bibnamefont {Jiang}}, \bibinfo {author} {\bibfnamefont {Y.}~\bibnamefont {Zhang}}, \bibinfo {author} {\bibfnamefont {S.~V.}\ \bibnamefont {Dubonos}}, \bibinfo {author} {\bibfnamefont {I.~V.}\ \bibnamefont {Grigorieva}},\ and\ \bibinfo {author} {\bibfnamefont {A.~A.}\ \bibnamefont {Firsov}},\ }\bibfield  {title} {\bibinfo {title} {Electric field effect in atomically thin carbon films},\ }\href {https://doi.org/10.1126/science.1102896} {\bibfield  {journal} {\bibinfo  {journal} {Science}\ }\textbf {\bibinfo {volume} {306}},\ \bibinfo {pages} {666} (\bibinfo {year} {2004})}\BibitemShut {NoStop}%
\bibitem [{\citenamefont {Novoselov}\ \emph {et~al.}(2005)\citenamefont {Novoselov}, \citenamefont {Geim}, \citenamefont {Morozov}, \citenamefont {Jiang}, \citenamefont {Katsnelson}, \citenamefont {Grigorieva}, \citenamefont {Dubonos},\ and\ \citenamefont {Firsov}}]{Novoselov_et_al_2005}%
  \BibitemOpen
  \bibfield  {author} {\bibinfo {author} {\bibfnamefont {K.~S.}\ \bibnamefont {Novoselov}}, \bibinfo {author} {\bibfnamefont {A.~K.}\ \bibnamefont {Geim}}, \bibinfo {author} {\bibfnamefont {S.~V.}\ \bibnamefont {Morozov}}, \bibinfo {author} {\bibfnamefont {D.}~\bibnamefont {Jiang}}, \bibinfo {author} {\bibfnamefont {M.~I.}\ \bibnamefont {Katsnelson}}, \bibinfo {author} {\bibfnamefont {I.~V.}\ \bibnamefont {Grigorieva}}, \bibinfo {author} {\bibfnamefont {S.~V.}\ \bibnamefont {Dubonos}},\ and\ \bibinfo {author} {\bibfnamefont {A.~A.}\ \bibnamefont {Firsov}},\ }\bibfield  {title} {\bibinfo {title} {Two-dimensional gas of massless {Dirac} fermions in graphene},\ }\href {https://doi.org/10.1038/nature04233} {\bibfield  {journal} {\bibinfo  {journal} {Nature}\ }\textbf {\bibinfo {volume} {438}},\ \bibinfo {pages} {197} (\bibinfo {year} {2005})}\BibitemShut {NoStop}%
\bibitem [{\citenamefont {Zhang}\ \emph {et~al.}(2005)\citenamefont {Zhang}, \citenamefont {Tan}, \citenamefont {Stormer},\ and\ \citenamefont {Kim}}]{Zhang_Kim_2005}%
  \BibitemOpen
  \bibfield  {author} {\bibinfo {author} {\bibfnamefont {Y.}~\bibnamefont {Zhang}}, \bibinfo {author} {\bibfnamefont {Y.-W.}\ \bibnamefont {Tan}}, \bibinfo {author} {\bibfnamefont {H.~L.}\ \bibnamefont {Stormer}},\ and\ \bibinfo {author} {\bibfnamefont {P.}~\bibnamefont {Kim}},\ }\bibfield  {title} {\bibinfo {title} {Experimental observation of the quantum {Hall} effect and berry's phase in graphene},\ }\href {https://doi.org/10.1038/nature04235} {\bibfield  {journal} {\bibinfo  {journal} {Nature}\ }\textbf {\bibinfo {volume} {438}},\ \bibinfo {pages} {201} (\bibinfo {year} {2005})}\BibitemShut {NoStop}%
\bibitem [{\citenamefont {Bolotin}\ \emph {et~al.}(2009)\citenamefont {Bolotin}, \citenamefont {Ghahari}, \citenamefont {Shulman}, \citenamefont {Stormer},\ and\ \citenamefont {Kim}}]{Bolotin_FQH2009}%
  \BibitemOpen
  \bibfield  {author} {\bibinfo {author} {\bibfnamefont {K.}~\bibnamefont {Bolotin}}, \bibinfo {author} {\bibfnamefont {F.}~\bibnamefont {Ghahari}}, \bibinfo {author} {\bibfnamefont {M.~D.}\ \bibnamefont {Shulman}}, \bibinfo {author} {\bibfnamefont {H.}~\bibnamefont {Stormer}},\ and\ \bibinfo {author} {\bibfnamefont {P.}~\bibnamefont {Kim}},\ }\bibfield  {title} {\bibinfo {title} {Observation of the fractional quantum {Hall} effect in graphene},\ }\href {https://doi.org/10.1038/nature08582} {\bibfield  {journal} {\bibinfo  {journal} {Nature}\ }\textbf {\bibinfo {volume} {462}},\ \bibinfo {pages} {196} (\bibinfo {year} {2009})}\BibitemShut {NoStop}%
\bibitem [{\citenamefont {Young}\ \emph {et~al.}(2012)\citenamefont {Young}, \citenamefont {Dean}, \citenamefont {Wang}, \citenamefont {Ren}, \citenamefont {Cadden-Zimansky}, \citenamefont {Watanabe}, \citenamefont {Taniguchi}, \citenamefont {Hone}, \citenamefont {Shepard},\ and\ \citenamefont {Kim}}]{young2012:nu0}%
  \BibitemOpen
  \bibfield  {author} {\bibinfo {author} {\bibfnamefont {A.~F.}\ \bibnamefont {Young}}, \bibinfo {author} {\bibfnamefont {C.~R.}\ \bibnamefont {Dean}}, \bibinfo {author} {\bibfnamefont {L.}~\bibnamefont {Wang}}, \bibinfo {author} {\bibfnamefont {H.}~\bibnamefont {Ren}}, \bibinfo {author} {\bibfnamefont {P.}~\bibnamefont {Cadden-Zimansky}}, \bibinfo {author} {\bibfnamefont {K.}~\bibnamefont {Watanabe}}, \bibinfo {author} {\bibfnamefont {T.}~\bibnamefont {Taniguchi}}, \bibinfo {author} {\bibfnamefont {J.}~\bibnamefont {Hone}}, \bibinfo {author} {\bibfnamefont {K.~L.}\ \bibnamefont {Shepard}},\ and\ \bibinfo {author} {\bibfnamefont {P.}~\bibnamefont {Kim}},\ }\bibfield  {title} {\bibinfo {title} {Spin and valley quantum {Hall} ferromagnetism in graphene},\ }\href {https://doi.org/10.1038/nphys2307} {\bibfield  {journal} {\bibinfo  {journal} {Nature Physics}\ }\textbf {\bibinfo {volume} {8}},\ \bibinfo {pages} {550–556} (\bibinfo {year} {2012})}\BibitemShut {NoStop}%
\bibitem [{\citenamefont {Feldman}\ \emph {et~al.}(2013)\citenamefont {Feldman}, \citenamefont {Levin}, \citenamefont {Krauss}, \citenamefont {Abanin}, \citenamefont {Halperin}, \citenamefont {Smet},\ and\ \citenamefont {Yacoby}}]{Phase_transition_vs_Bperp_Yacoby2013}%
  \BibitemOpen
  \bibfield  {author} {\bibinfo {author} {\bibfnamefont {B.~E.}\ \bibnamefont {Feldman}}, \bibinfo {author} {\bibfnamefont {A.~J.}\ \bibnamefont {Levin}}, \bibinfo {author} {\bibfnamefont {B.}~\bibnamefont {Krauss}}, \bibinfo {author} {\bibfnamefont {D.~A.}\ \bibnamefont {Abanin}}, \bibinfo {author} {\bibfnamefont {B.~I.}\ \bibnamefont {Halperin}}, \bibinfo {author} {\bibfnamefont {J.~H.}\ \bibnamefont {Smet}},\ and\ \bibinfo {author} {\bibfnamefont {A.}~\bibnamefont {Yacoby}},\ }\bibfield  {title} {\bibinfo {title} {Fractional quantum {Hall} phase transitions and four-flux states in graphene},\ }\href {https://doi.org/10.1103/PhysRevLett.111.076802} {\bibfield  {journal} {\bibinfo  {journal} {Phys. Rev. Lett.}\ }\textbf {\bibinfo {volume} {111}},\ \bibinfo {pages} {076802} (\bibinfo {year} {2013})}\BibitemShut {NoStop}%
\bibitem [{\citenamefont {Young}\ \emph {et~al.}(2014)\citenamefont {Young}, \citenamefont {Sanchez-Yamagishi}, \citenamefont {Hunt}, \citenamefont {Choi}, \citenamefont {Watanabe}, \citenamefont {Taniguchi}, \citenamefont {Ashoori},\ and\ \citenamefont {Jarillo-Herrero}}]{young2014:nu0}%
  \BibitemOpen
  \bibfield  {author} {\bibinfo {author} {\bibfnamefont {A.~F.}\ \bibnamefont {Young}}, \bibinfo {author} {\bibfnamefont {J.~D.}\ \bibnamefont {Sanchez-Yamagishi}}, \bibinfo {author} {\bibfnamefont {B.}~\bibnamefont {Hunt}}, \bibinfo {author} {\bibfnamefont {S.~H.}\ \bibnamefont {Choi}}, \bibinfo {author} {\bibfnamefont {K.}~\bibnamefont {Watanabe}}, \bibinfo {author} {\bibfnamefont {T.}~\bibnamefont {Taniguchi}}, \bibinfo {author} {\bibfnamefont {R.~C.}\ \bibnamefont {Ashoori}},\ and\ \bibinfo {author} {\bibfnamefont {P.}~\bibnamefont {Jarillo-Herrero}},\ }\bibfield  {title} {\bibinfo {title} {Tunable symmetry breaking and helical edge transport in a graphene quantum spin {Hall} state},\ }\href {https://doi.org/10.1038/nature12800} {\bibfield  {journal} {\bibinfo  {journal} {Nature}\ }\textbf {\bibinfo {volume} {505}},\ \bibinfo {pages} {528} (\bibinfo {year} {2014})}\BibitemShut {NoStop}%
\bibitem [{\citenamefont {Amet}\ \emph {et~al.}(2015)\citenamefont {Amet}, \citenamefont {Bestwick}, \citenamefont {Williams}, \citenamefont {Balicas}, \citenamefont {Watanabe}, \citenamefont {Taniguchi},\ and\ \citenamefont {Goldhaber-Gordon}}]{GoldhaberGordon2015}%
  \BibitemOpen
  \bibfield  {author} {\bibinfo {author} {\bibfnamefont {F.}~\bibnamefont {Amet}}, \bibinfo {author} {\bibfnamefont {A.~J.}\ \bibnamefont {Bestwick}}, \bibinfo {author} {\bibfnamefont {J.~R.}\ \bibnamefont {Williams}}, \bibinfo {author} {\bibfnamefont {L.}~\bibnamefont {Balicas}}, \bibinfo {author} {\bibfnamefont {K.}~\bibnamefont {Watanabe}}, \bibinfo {author} {\bibfnamefont {T.}~\bibnamefont {Taniguchi}},\ and\ \bibinfo {author} {\bibfnamefont {D.}~\bibnamefont {Goldhaber-Gordon}},\ }\bibfield  {title} {\bibinfo {title} {Composite fermions and broken symmetries in graphene},\ }\href {https://doi.org/10.1038/ncomms6838} {\bibfield  {journal} {\bibinfo  {journal} {Nat. Commun.}\ }\textbf {\bibinfo {volume} {6}},\ \bibinfo {pages} {5838} (\bibinfo {year} {2015})}\BibitemShut {NoStop}%
\bibitem [{\citenamefont {Zibrov}\ \emph {et~al.}(2018)\citenamefont {Zibrov}, \citenamefont {Spanton}, \citenamefont {Zhou}, \citenamefont {Kometter}, \citenamefont {Taniguchi}, \citenamefont {Watanabe},\ and\ \citenamefont {Young}}]{Even_Den_Young2018}%
  \BibitemOpen
  \bibfield  {author} {\bibinfo {author} {\bibfnamefont {A.~A.}\ \bibnamefont {Zibrov}}, \bibinfo {author} {\bibfnamefont {E.~M.}\ \bibnamefont {Spanton}}, \bibinfo {author} {\bibfnamefont {H.}~\bibnamefont {Zhou}}, \bibinfo {author} {\bibfnamefont {C.}~\bibnamefont {Kometter}}, \bibinfo {author} {\bibfnamefont {T.}~\bibnamefont {Taniguchi}}, \bibinfo {author} {\bibfnamefont {K.}~\bibnamefont {Watanabe}},\ and\ \bibinfo {author} {\bibfnamefont {A.~F.}\ \bibnamefont {Young}},\ }\bibfield  {title} {\bibinfo {title} {Even-denominator fractional quantum {Hall} states at an isospin transition in monolayer graphene},\ }\href {https://doi.org/10.1038/s41567-018-0190-0} {\bibfield  {journal} {\bibinfo  {journal} {Nature Physics}\ }\textbf {\bibinfo {volume} {14}},\ \bibinfo {pages} {930} (\bibinfo {year} {2018})}\BibitemShut {NoStop}%
\bibitem [{\citenamefont {Kim}\ \emph {et~al.}(2019)\citenamefont {Kim}, \citenamefont {Balram}, \citenamefont {Taniguchi}, \citenamefont {Watanabe}, \citenamefont {Jain},\ and\ \citenamefont {Smet}}]{Kim19}%
  \BibitemOpen
  \bibfield  {author} {\bibinfo {author} {\bibfnamefont {Y.}~\bibnamefont {Kim}}, \bibinfo {author} {\bibfnamefont {A.~C.}\ \bibnamefont {Balram}}, \bibinfo {author} {\bibfnamefont {T.}~\bibnamefont {Taniguchi}}, \bibinfo {author} {\bibfnamefont {K.}~\bibnamefont {Watanabe}}, \bibinfo {author} {\bibfnamefont {J.~K.}\ \bibnamefont {Jain}},\ and\ \bibinfo {author} {\bibfnamefont {J.~H.}\ \bibnamefont {Smet}},\ }\bibfield  {title} {\bibinfo {title} {Even denominator fractional quantum {Hall} states in higher {Landau} levels of graphene},\ }\href {https://doi.org/10.1038/s41567-018-0355-x} {\bibfield  {journal} {\bibinfo  {journal} {Nature Physics}\ }\textbf {\bibinfo {volume} {15}},\ \bibinfo {pages} {154} (\bibinfo {year} {2019})}\BibitemShut {NoStop}%
\bibitem [{\citenamefont {Alicea}\ and\ \citenamefont {Fisher}(2006)}]{alicea2006:gqhe}%
  \BibitemOpen
  \bibfield  {author} {\bibinfo {author} {\bibfnamefont {J.}~\bibnamefont {Alicea}}\ and\ \bibinfo {author} {\bibfnamefont {M.~P.~A.}\ \bibnamefont {Fisher}},\ }\bibfield  {title} {\bibinfo {title} {Graphene integer quantum {Hall} effect in the ferromagnetic and paramagnetic regimes},\ }\href {https://doi.org/10.1103/PhysRevB.74.075422} {\bibfield  {journal} {\bibinfo  {journal} {Phys. Rev. B}\ }\textbf {\bibinfo {volume} {74}},\ \bibinfo {pages} {075422} (\bibinfo {year} {2006})}\BibitemShut {NoStop}%
\bibitem [{\citenamefont {Abanin}\ \emph {et~al.}(2006)\citenamefont {Abanin}, \citenamefont {Lee},\ and\ \citenamefont {Levitov}}]{abanin2006:nu0}%
  \BibitemOpen
  \bibfield  {author} {\bibinfo {author} {\bibfnamefont {D.~A.}\ \bibnamefont {Abanin}}, \bibinfo {author} {\bibfnamefont {P.~A.}\ \bibnamefont {Lee}},\ and\ \bibinfo {author} {\bibfnamefont {L.~S.}\ \bibnamefont {Levitov}},\ }\bibfield  {title} {\bibinfo {title} {Spin-filtered edge states and quantum {Hall} effect in graphene},\ }\href {https://doi.org/10.1103/PhysRevLett.96.176803} {\bibfield  {journal} {\bibinfo  {journal} {Phys. Rev. Lett.}\ }\textbf {\bibinfo {volume} {96}},\ \bibinfo {pages} {176803} (\bibinfo {year} {2006})}\BibitemShut {NoStop}%
\bibitem [{\citenamefont {Fertig}\ and\ \citenamefont {Brey}(2006)}]{brey2006:nu0}%
  \BibitemOpen
  \bibfield  {author} {\bibinfo {author} {\bibfnamefont {H.~A.}\ \bibnamefont {Fertig}}\ and\ \bibinfo {author} {\bibfnamefont {L.}~\bibnamefont {Brey}},\ }\bibfield  {title} {\bibinfo {title} {Luttinger liquid at the edge of undoped graphene in a strong magnetic field},\ }\href {https://doi.org/10.1103/PhysRevLett.97.116805} {\bibfield  {journal} {\bibinfo  {journal} {Phys. Rev. Lett.}\ }\textbf {\bibinfo {volume} {97}},\ \bibinfo {pages} {116805} (\bibinfo {year} {2006})}\BibitemShut {NoStop}%
\bibitem [{\citenamefont {Herbut}(2007{\natexlab{a}})}]{Herbut1}%
  \BibitemOpen
  \bibfield  {author} {\bibinfo {author} {\bibfnamefont {I.~F.}\ \bibnamefont {Herbut}},\ }\bibfield  {title} {\bibinfo {title} {Theory of integer quantum {Hall} effect in graphene},\ }\href {https://doi.org/10.1103/PhysRevB.75.165411} {\bibfield  {journal} {\bibinfo  {journal} {Phys. Rev. B}\ }\textbf {\bibinfo {volume} {75}},\ \bibinfo {pages} {165411} (\bibinfo {year} {2007}{\natexlab{a}})}\BibitemShut {NoStop}%
\bibitem [{\citenamefont {Herbut}(2007{\natexlab{b}})}]{Herbut2}%
  \BibitemOpen
  \bibfield  {author} {\bibinfo {author} {\bibfnamefont {I.~F.}\ \bibnamefont {Herbut}},\ }\bibfield  {title} {\bibinfo {title} {{S}{O}(3) symmetry between {N\'eel} and ferromagnetic order parameters for graphene in a magnetic field},\ }\href {https://doi.org/10.1103/PhysRevB.76.085432} {\bibfield  {journal} {\bibinfo  {journal} {Phys. Rev. B}\ }\textbf {\bibinfo {volume} {76}},\ \bibinfo {pages} {085432} (\bibinfo {year} {2007}{\natexlab{b}})}\BibitemShut {NoStop}%
\bibitem [{\citenamefont {Yang}\ \emph {et~al.}(2006)\citenamefont {Yang}, \citenamefont {Das~Sarma},\ and\ \citenamefont {MacDonald}}]{KYang_SU4_Skyrmion_2006}%
  \BibitemOpen
  \bibfield  {author} {\bibinfo {author} {\bibfnamefont {K.}~\bibnamefont {Yang}}, \bibinfo {author} {\bibfnamefont {S.}~\bibnamefont {Das~Sarma}},\ and\ \bibinfo {author} {\bibfnamefont {A.~H.}\ \bibnamefont {MacDonald}},\ }\bibfield  {title} {\bibinfo {title} {Collective modes and skyrmion excitations in graphene ${S}{U}(4)$ quantum {Hall} ferromagnets},\ }\href {https://doi.org/10.1103/PhysRevB.74.075423} {\bibfield  {journal} {\bibinfo  {journal} {Phys. Rev. B}\ }\textbf {\bibinfo {volume} {74}},\ \bibinfo {pages} {075423} (\bibinfo {year} {2006})}\BibitemShut {NoStop}%
\bibitem [{\citenamefont {Kharitonov}(2012)}]{kharitonov2012:nu0}%
  \BibitemOpen
  \bibfield  {author} {\bibinfo {author} {\bibfnamefont {M.}~\bibnamefont {Kharitonov}},\ }\bibfield  {title} {\bibinfo {title} {Phase diagram for the $\ensuremath{\nu}=0$ quantum {Hall} state in monolayer graphene},\ }\href {https://doi.org/10.1103/PhysRevB.85.155439} {\bibfield  {journal} {\bibinfo  {journal} {Phys. Rev. B}\ }\textbf {\bibinfo {volume} {85}},\ \bibinfo {pages} {155439} (\bibinfo {year} {2012})}\BibitemShut {NoStop}%
\bibitem [{\citenamefont {Das}\ \emph {et~al.}(2022)\citenamefont {Das}, \citenamefont {Kaul},\ and\ \citenamefont {Murthy}}]{Das_Kaul_Murthy_2022}%
  \BibitemOpen
  \bibfield  {author} {\bibinfo {author} {\bibfnamefont {A.}~\bibnamefont {Das}}, \bibinfo {author} {\bibfnamefont {R.~K.}\ \bibnamefont {Kaul}},\ and\ \bibinfo {author} {\bibfnamefont {G.}~\bibnamefont {Murthy}},\ }\bibfield  {title} {\bibinfo {title} {Coexistence of canted antiferromagnetism and bond order in $\ensuremath{\nu}=0$ graphene},\ }\href {https://doi.org/10.1103/PhysRevLett.128.106803} {\bibfield  {journal} {\bibinfo  {journal} {Phys. Rev. Lett.}\ }\textbf {\bibinfo {volume} {128}},\ \bibinfo {pages} {106803} (\bibinfo {year} {2022})}\BibitemShut {NoStop}%
\bibitem [{\citenamefont {De}\ \emph {et~al.}(2023)\citenamefont {De}, \citenamefont {Das}, \citenamefont {Rao}, \citenamefont {Kaul},\ and\ \citenamefont {Murthy}}]{De_etal_Murthy2022}%
  \BibitemOpen
  \bibfield  {author} {\bibinfo {author} {\bibfnamefont {S.~J.}\ \bibnamefont {De}}, \bibinfo {author} {\bibfnamefont {A.}~\bibnamefont {Das}}, \bibinfo {author} {\bibfnamefont {S.}~\bibnamefont {Rao}}, \bibinfo {author} {\bibfnamefont {R.~K.}\ \bibnamefont {Kaul}},\ and\ \bibinfo {author} {\bibfnamefont {G.}~\bibnamefont {Murthy}},\ }\bibfield  {title} {\bibinfo {title} {Global phase diagram of charge-neutral graphene in the quantum {Hall} regime for generic interactions},\ }\href {https://doi.org/10.1103/PhysRevB.107.125422} {\bibfield  {journal} {\bibinfo  {journal} {Phys. Rev. B}\ }\textbf {\bibinfo {volume} {107}},\ \bibinfo {pages} {125422} (\bibinfo {year} {2023})}\BibitemShut {NoStop}%
\bibitem [{\citenamefont {Stefanidis}\ and\ \citenamefont {Villadiego}(2023)}]{Stefanidis_Sodemann2022}%
  \BibitemOpen
  \bibfield  {author} {\bibinfo {author} {\bibfnamefont {N.}~\bibnamefont {Stefanidis}}\ and\ \bibinfo {author} {\bibfnamefont {I.~S.}\ \bibnamefont {Villadiego}},\ }\bibfield  {title} {\bibinfo {title} {Competing spin-valley entangled and broken symmetry states in the $n=1$ {Landau} level of graphene},\ }\href {https://doi.org/10.1103/PhysRevB.107.045132} {\bibfield  {journal} {\bibinfo  {journal} {Phys. Rev. B}\ }\textbf {\bibinfo {volume} {107}},\ \bibinfo {pages} {045132} (\bibinfo {year} {2023})}\BibitemShut {NoStop}%
\bibitem [{\citenamefont {Sodemann}\ and\ \citenamefont {MacDonald}(2014)}]{Sodemann_MacDonald_2014}%
  \BibitemOpen
  \bibfield  {author} {\bibinfo {author} {\bibfnamefont {I.}~\bibnamefont {Sodemann}}\ and\ \bibinfo {author} {\bibfnamefont {A.~H.}\ \bibnamefont {MacDonald}},\ }\bibfield  {title} {\bibinfo {title} {Broken {S}{U}(4) symmetry and the fractional quantum {Hall} effect in graphene},\ }\href {https://doi.org/10.1103/PhysRevLett.112.126804} {\bibfield  {journal} {\bibinfo  {journal} {Phys. Rev. Lett.}\ }\textbf {\bibinfo {volume} {112}},\ \bibinfo {pages} {126804} (\bibinfo {year} {2014})}\BibitemShut {NoStop}%
\bibitem [{\citenamefont {An}\ \emph {et~al.}(2024)\citenamefont {An}, \citenamefont {Balram},\ and\ \citenamefont {Murthy}}]{Jincheng2024}%
  \BibitemOpen
  \bibfield  {author} {\bibinfo {author} {\bibfnamefont {J.}~\bibnamefont {An}}, \bibinfo {author} {\bibfnamefont {A.~C.}\ \bibnamefont {Balram}},\ and\ \bibinfo {author} {\bibfnamefont {G.}~\bibnamefont {Murthy}},\ }\bibfield  {title} {\bibinfo {title} {Magnetic and lattice ordered fractional quantum {Hall} phases in graphene},\ }\href {https://doi.org/10.1103/PhysRevB.110.L081103} {\bibfield  {journal} {\bibinfo  {journal} {Phys. Rev. B}\ }\textbf {\bibinfo {volume} {110}},\ \bibinfo {pages} {L081103} (\bibinfo {year} {2024})}\BibitemShut {NoStop}%
\bibitem [{\citenamefont {Wei}\ \emph {et~al.}(2018)\citenamefont {Wei}, \citenamefont {van~der Sar}, \citenamefont {Lee}, \citenamefont {Watanabe}, \citenamefont {Taniguchi}, \citenamefont {Halperin},\ and\ \citenamefont {Yacoby}}]{Magnon_transport_Yacoby_2018}%
  \BibitemOpen
  \bibfield  {author} {\bibinfo {author} {\bibfnamefont {D.~S.}\ \bibnamefont {Wei}}, \bibinfo {author} {\bibfnamefont {T.}~\bibnamefont {van~der Sar}}, \bibinfo {author} {\bibfnamefont {S.~H.}\ \bibnamefont {Lee}}, \bibinfo {author} {\bibfnamefont {K.}~\bibnamefont {Watanabe}}, \bibinfo {author} {\bibfnamefont {T.}~\bibnamefont {Taniguchi}}, \bibinfo {author} {\bibfnamefont {B.~I.}\ \bibnamefont {Halperin}},\ and\ \bibinfo {author} {\bibfnamefont {A.}~\bibnamefont {Yacoby}},\ }\bibfield  {title} {\bibinfo {title} {Electrical generation and detection of spin waves in a quantum {Hall} ferromagnet},\ }\href {https://doi.org/10.1126/science.aar4061} {\bibfield  {journal} {\bibinfo  {journal} {Science}\ }\textbf {\bibinfo {volume} {362}},\ \bibinfo {pages} {229–233} (\bibinfo {year} {2018})}\BibitemShut {NoStop}%
\bibitem [{\citenamefont {Liu}\ \emph {et~al.}(2022)\citenamefont {Liu}, \citenamefont {Farahi}, \citenamefont {Chiu}, \citenamefont {Papic}, \citenamefont {Watanabe}, \citenamefont {Taniguchi}, \citenamefont {Zaletel},\ and\ \citenamefont {Yazdani}}]{STM_Yazdani2021visualizing}%
  \BibitemOpen
  \bibfield  {author} {\bibinfo {author} {\bibfnamefont {X.}~\bibnamefont {Liu}}, \bibinfo {author} {\bibfnamefont {G.}~\bibnamefont {Farahi}}, \bibinfo {author} {\bibfnamefont {C.-L.}\ \bibnamefont {Chiu}}, \bibinfo {author} {\bibfnamefont {Z.}~\bibnamefont {Papic}}, \bibinfo {author} {\bibfnamefont {K.}~\bibnamefont {Watanabe}}, \bibinfo {author} {\bibfnamefont {T.}~\bibnamefont {Taniguchi}}, \bibinfo {author} {\bibfnamefont {M.~P.}\ \bibnamefont {Zaletel}},\ and\ \bibinfo {author} {\bibfnamefont {A.}~\bibnamefont {Yazdani}},\ }\bibfield  {title} {\bibinfo {title} {Visualizing broken symmetry and topological defects in a quantum {Hall} ferromagnet},\ }\href {https://doi.org/10.1126/science.abm3770} {\bibfield  {journal} {\bibinfo  {journal} {Science}\ }\textbf {\bibinfo {volume} {375}},\ \bibinfo {pages} {321–326} (\bibinfo {year} {2022})}\BibitemShut {NoStop}%
\bibitem [{\citenamefont {Coissard}\ \emph {et~al.}(2022)\citenamefont {Coissard}, \citenamefont {Wander}, \citenamefont {Vignaud}, \citenamefont {Grushin}, \citenamefont {Repellin}, \citenamefont {Watanabe}, \citenamefont {Taniguchi}, \citenamefont {Gay}, \citenamefont {Winkelmann}, \citenamefont {Courtois}, \citenamefont {Sellier},\ and\ \citenamefont {Sac{\'{e}}p{\'{e}}}}]{STM_Coissard_2022}%
  \BibitemOpen
  \bibfield  {author} {\bibinfo {author} {\bibfnamefont {A.}~\bibnamefont {Coissard}}, \bibinfo {author} {\bibfnamefont {D.}~\bibnamefont {Wander}}, \bibinfo {author} {\bibfnamefont {H.}~\bibnamefont {Vignaud}}, \bibinfo {author} {\bibfnamefont {A.~G.}\ \bibnamefont {Grushin}}, \bibinfo {author} {\bibfnamefont {C.}~\bibnamefont {Repellin}}, \bibinfo {author} {\bibfnamefont {K.}~\bibnamefont {Watanabe}}, \bibinfo {author} {\bibfnamefont {T.}~\bibnamefont {Taniguchi}}, \bibinfo {author} {\bibfnamefont {F.}~\bibnamefont {Gay}}, \bibinfo {author} {\bibfnamefont {C.~B.}\ \bibnamefont {Winkelmann}}, \bibinfo {author} {\bibfnamefont {H.}~\bibnamefont {Courtois}}, \bibinfo {author} {\bibfnamefont {H.}~\bibnamefont {Sellier}},\ and\ \bibinfo {author} {\bibfnamefont {B.}~\bibnamefont {Sac{\'{e}}p{\'{e}}}},\ }\bibfield  {title} {\bibinfo {title} {Imaging tunable quantum {Hall} broken-symmetry orders in graphene},\ }\href {https://doi.org/10.1038/s41586-022-04513-7} {\bibfield  {journal} {\bibinfo  {journal} {Nature}\
  }\textbf {\bibinfo {volume} {605}},\ \bibinfo {pages} {51} (\bibinfo {year} {2022})}\BibitemShut {NoStop}%
\bibitem [{\citenamefont {Balram}\ \emph {et~al.}(2015{\natexlab{a}})\citenamefont {Balram}, \citenamefont {T\"oke}, \citenamefont {W\'ojs},\ and\ \citenamefont {Jain}}]{Balram15a}%
  \BibitemOpen
  \bibfield  {author} {\bibinfo {author} {\bibfnamefont {A.~C.}\ \bibnamefont {Balram}}, \bibinfo {author} {\bibfnamefont {C.}~\bibnamefont {T\"oke}}, \bibinfo {author} {\bibfnamefont {A.}~\bibnamefont {W\'ojs}},\ and\ \bibinfo {author} {\bibfnamefont {J.~K.}\ \bibnamefont {Jain}},\ }\bibfield  {title} {\bibinfo {title} {Fractional quantum {Hall} effect in graphene: Quantitative comparison between theory and experiment},\ }\href {https://doi.org/10.1103/PhysRevB.92.075410} {\bibfield  {journal} {\bibinfo  {journal} {Phys. Rev. B}\ }\textbf {\bibinfo {volume} {92}},\ \bibinfo {pages} {075410} (\bibinfo {year} {2015}{\natexlab{a}})}\BibitemShut {NoStop}%
\bibitem [{\citenamefont {Balram}\ \emph {et~al.}(2015{\natexlab{b}})\citenamefont {Balram}, \citenamefont {T\ifmmode~\mbox{\H{o}}\else \H{o}\fi{}ke}, \citenamefont {W\'ojs},\ and\ \citenamefont {Jain}}]{Balram15c}%
  \BibitemOpen
  \bibfield  {author} {\bibinfo {author} {\bibfnamefont {A.~C.}\ \bibnamefont {Balram}}, \bibinfo {author} {\bibfnamefont {C.}~\bibnamefont {T\ifmmode~\mbox{\H{o}}\else \H{o}\fi{}ke}}, \bibinfo {author} {\bibfnamefont {A.}~\bibnamefont {W\'ojs}},\ and\ \bibinfo {author} {\bibfnamefont {J.~K.}\ \bibnamefont {Jain}},\ }\bibfield  {title} {\bibinfo {title} {Spontaneous polarization of composite fermions in the $n=1$ {Landau} level of graphene},\ }\href {https://doi.org/10.1103/PhysRevB.92.205120} {\bibfield  {journal} {\bibinfo  {journal} {Phys. Rev. B}\ }\textbf {\bibinfo {volume} {92}},\ \bibinfo {pages} {205120} (\bibinfo {year} {2015}{\natexlab{b}})}\BibitemShut {NoStop}%
\bibitem [{\citenamefont {Jain}(1989)}]{Jain89}%
  \BibitemOpen
  \bibfield  {author} {\bibinfo {author} {\bibfnamefont {J.~K.}\ \bibnamefont {Jain}},\ }\bibfield  {title} {\bibinfo {title} {Composite-fermion approach for the fractional quantum {Hall} effect},\ }\href {https://doi.org/10.1103/PhysRevLett.63.199} {\bibfield  {journal} {\bibinfo  {journal} {Phys. Rev. Lett.}\ }\textbf {\bibinfo {volume} {63}},\ \bibinfo {pages} {199} (\bibinfo {year} {1989})}\BibitemShut {NoStop}%
\bibitem [{\citenamefont {Abanin}\ \emph {et~al.}(2013)\citenamefont {Abanin}, \citenamefont {Feldman}, \citenamefont {Yacoby},\ and\ \citenamefont {Halperin}}]{Abanin_Halperin2013}%
  \BibitemOpen
  \bibfield  {author} {\bibinfo {author} {\bibfnamefont {D.~A.}\ \bibnamefont {Abanin}}, \bibinfo {author} {\bibfnamefont {B.~E.}\ \bibnamefont {Feldman}}, \bibinfo {author} {\bibfnamefont {A.}~\bibnamefont {Yacoby}},\ and\ \bibinfo {author} {\bibfnamefont {B.~I.}\ \bibnamefont {Halperin}},\ }\bibfield  {title} {\bibinfo {title} {Fractional and integer quantum {Hall} effects in the zeroth {Landau} level in graphene},\ }\href {https://doi.org/10.1103/PhysRevB.88.115407} {\bibfield  {journal} {\bibinfo  {journal} {Phys. Rev. B}\ }\textbf {\bibinfo {volume} {88}},\ \bibinfo {pages} {115407} (\bibinfo {year} {2013})}\BibitemShut {NoStop}%
\bibitem [{\citenamefont {Huang}\ \emph {et~al.}(2022)\citenamefont {Huang}, \citenamefont {Fu}, \citenamefont {Hickey}, \citenamefont {Alem}, \citenamefont {Lin}, \citenamefont {Watanabe}, \citenamefont {Taniguchi},\ and\ \citenamefont {Zhu}}]{HUang21}%
  \BibitemOpen
  \bibfield  {author} {\bibinfo {author} {\bibfnamefont {K.}~\bibnamefont {Huang}}, \bibinfo {author} {\bibfnamefont {H.}~\bibnamefont {Fu}}, \bibinfo {author} {\bibfnamefont {D.~R.}\ \bibnamefont {Hickey}}, \bibinfo {author} {\bibfnamefont {N.}~\bibnamefont {Alem}}, \bibinfo {author} {\bibfnamefont {X.}~\bibnamefont {Lin}}, \bibinfo {author} {\bibfnamefont {K.}~\bibnamefont {Watanabe}}, \bibinfo {author} {\bibfnamefont {T.}~\bibnamefont {Taniguchi}},\ and\ \bibinfo {author} {\bibfnamefont {J.}~\bibnamefont {Zhu}},\ }\bibfield  {title} {\bibinfo {title} {Valley isospin controlled fractional quantum {Hall} states in bilayer graphene},\ }\href {https://doi.org/10.1103/PhysRevX.12.031019} {\bibfield  {journal} {\bibinfo  {journal} {Phys. Rev. X}\ }\textbf {\bibinfo {volume} {12}},\ \bibinfo {pages} {031019} (\bibinfo {year} {2022})}\BibitemShut {NoStop}%
\bibitem [{\citenamefont {An}\ \emph {et~al.}(2025)\citenamefont {An}, \citenamefont {Balram}, \citenamefont {Khanna},\ and\ \citenamefont {Murthy}}]{an2024fractional}%
  \BibitemOpen
  \bibfield  {author} {\bibinfo {author} {\bibfnamefont {J.}~\bibnamefont {An}}, \bibinfo {author} {\bibfnamefont {A.~C.}\ \bibnamefont {Balram}}, \bibinfo {author} {\bibfnamefont {U.}~\bibnamefont {Khanna}},\ and\ \bibinfo {author} {\bibfnamefont {G.}~\bibnamefont {Murthy}},\ }\bibfield  {title} {\bibinfo {title} {Fractional quantum {Hall} coexistence phases in higher {Landau} levels of graphene},\ }\href {https://doi.org/10.1103/PhysRevB.111.045110} {\bibfield  {journal} {\bibinfo  {journal} {Phys. Rev. B}\ }\textbf {\bibinfo {volume} {111}},\ \bibinfo {pages} {045110} (\bibinfo {year} {2025})}\BibitemShut {NoStop}%
\bibitem [{\citenamefont {Aleiner}\ \emph {et~al.}(2007)\citenamefont {Aleiner}, \citenamefont {Kharzeev},\ and\ \citenamefont {Tsvelik}}]{RG1_Graphene_Aleiner_2007}%
  \BibitemOpen
  \bibfield  {author} {\bibinfo {author} {\bibfnamefont {I.~L.}\ \bibnamefont {Aleiner}}, \bibinfo {author} {\bibfnamefont {D.~E.}\ \bibnamefont {Kharzeev}},\ and\ \bibinfo {author} {\bibfnamefont {A.~M.}\ \bibnamefont {Tsvelik}},\ }\bibfield  {title} {\bibinfo {title} {Spontaneous symmetry breaking in graphene subjected to an in-plane magnetic field},\ }\href {https://doi.org/10.1103/PhysRevB.76.195415} {\bibfield  {journal} {\bibinfo  {journal} {Phys. Rev. B}\ }\textbf {\bibinfo {volume} {76}},\ \bibinfo {pages} {195415} (\bibinfo {year} {2007})}\BibitemShut {NoStop}%
\bibitem [{\citenamefont {Basko}\ and\ \citenamefont {Aleiner}(2008)}]{RG2_Graphene_Aleiner_2008}%
  \BibitemOpen
  \bibfield  {author} {\bibinfo {author} {\bibfnamefont {D.~M.}\ \bibnamefont {Basko}}\ and\ \bibinfo {author} {\bibfnamefont {I.~L.}\ \bibnamefont {Aleiner}},\ }\bibfield  {title} {\bibinfo {title} {Interplay of coulomb and electron-phonon interactions in graphene},\ }\href {https://doi.org/10.1103/PhysRevB.77.041409} {\bibfield  {journal} {\bibinfo  {journal} {Phys. Rev. B}\ }\textbf {\bibinfo {volume} {77}},\ \bibinfo {pages} {041409} (\bibinfo {year} {2008})}\BibitemShut {NoStop}%
\bibitem [{\citenamefont {Haldane}(1983)}]{Haldane_Pseudopot1983}%
  \BibitemOpen
  \bibfield  {author} {\bibinfo {author} {\bibfnamefont {F.~D.~M.}\ \bibnamefont {Haldane}},\ }\bibfield  {title} {\bibinfo {title} {Fractional quantization of the {Hall} effect: A hierarchy of incompressible quantum fluid states},\ }\href {https://doi.org/10.1103/PhysRevLett.51.605} {\bibfield  {journal} {\bibinfo  {journal} {Phys. Rev. Lett.}\ }\textbf {\bibinfo {volume} {51}},\ \bibinfo {pages} {605} (\bibinfo {year} {1983})}\BibitemShut {NoStop}%
\bibitem [{SM()}]{SM}%
  \BibitemOpen
  \href@noop {} {}\bibinfo {note} {See Supplemental Material for the technical details about the results presented in the main text, including (i) the variational energy functional for fractional quantum Hall states at filling $(1,\nu_F)$ in the $\bn{=}0$ and $\bn{=}1$ Landau levels of graphene, (ii) the spinor ansatz used for energy minimization, along with various order parameters that distinguish all possible phases, (iii) the spinors for all fractional quantum Hall phases at filling $(1,\nu_F)$ that we have considered, (iv) the formalism to evaluate the transport gaps of fractional quantum Hall states at filling $(1,\nu_F)$, and (v) reasons why the most obvious candidate for producing the anomalous $E_Z$ dependence in the 1LLs~\cite{GoldhaberGordon2015}, namely, the V/SAF state, cannot possibly do so.}\BibitemShut {Stop}%
\bibitem [{\citenamefont {Morf}\ \emph {et~al.}(2002)\citenamefont {Morf}, \citenamefont {d'Ambrumenil},\ and\ \citenamefont {Das~Sarma}}]{Morf02}%
  \BibitemOpen
  \bibfield  {author} {\bibinfo {author} {\bibfnamefont {R.~H.}\ \bibnamefont {Morf}}, \bibinfo {author} {\bibfnamefont {N.}~\bibnamefont {d'Ambrumenil}},\ and\ \bibinfo {author} {\bibfnamefont {S.}~\bibnamefont {Das~Sarma}},\ }\bibfield  {title} {\bibinfo {title} {Excitation gaps in fractional quantum {Hall} states: An exact diagonalization study},\ }\href {https://doi.org/10.1103/PhysRevB.66.075408} {\bibfield  {journal} {\bibinfo  {journal} {Phys. Rev. B}\ }\textbf {\bibinfo {volume} {66}},\ \bibinfo {pages} {075408} (\bibinfo {year} {2002})}\BibitemShut {NoStop}%
\bibitem [{\citenamefont {Balram}\ and\ \citenamefont {W\'ojs}(2020)}]{Balram20b}%
  \BibitemOpen
  \bibfield  {author} {\bibinfo {author} {\bibfnamefont {A.~C.}\ \bibnamefont {Balram}}\ and\ \bibinfo {author} {\bibfnamefont {A.}~\bibnamefont {W\'ojs}},\ }\bibfield  {title} {\bibinfo {title} {Fractional quantum {Hall} effect at $\ensuremath{\nu}=2+4/9$},\ }\href {https://doi.org/10.1103/PhysRevResearch.2.032035} {\bibfield  {journal} {\bibinfo  {journal} {Phys. Rev. Research}\ }\textbf {\bibinfo {volume} {2}},\ \bibinfo {pages} {032035} (\bibinfo {year} {2020})}\BibitemShut {NoStop}%
\bibitem [{\citenamefont {Wu}\ \emph {et~al.}(1993)\citenamefont {Wu}, \citenamefont {Dev},\ and\ \citenamefont {Jain}}]{Wu93}%
  \BibitemOpen
  \bibfield  {author} {\bibinfo {author} {\bibfnamefont {X.~G.}\ \bibnamefont {Wu}}, \bibinfo {author} {\bibfnamefont {G.}~\bibnamefont {Dev}},\ and\ \bibinfo {author} {\bibfnamefont {J.~K.}\ \bibnamefont {Jain}},\ }\bibfield  {title} {\bibinfo {title} {Mixed-spin incompressible states in the fractional quantum {Hall} effect},\ }\href {https://doi.org/10.1103/PhysRevLett.71.153} {\bibfield  {journal} {\bibinfo  {journal} {Phys. Rev. Lett.}\ }\textbf {\bibinfo {volume} {71}},\ \bibinfo {pages} {153} (\bibinfo {year} {1993})}\BibitemShut {NoStop}%
\bibitem [{\citenamefont {Hunt}\ \emph {et~al.}(2016)\citenamefont {Hunt}, \citenamefont {Li}, \citenamefont {Zibrov}, \citenamefont {Wang}, \citenamefont {Taniguchi}, \citenamefont {Watanabe}, \citenamefont {Hone}, \citenamefont {Dean}, \citenamefont {Zaletel}, \citenamefont {Ashoori}, \citenamefont {Young},\ and\ \citenamefont {Young}}]{Hunt2016DirectMO}%
  \BibitemOpen
  \bibfield  {author} {\bibinfo {author} {\bibfnamefont {B.}~\bibnamefont {Hunt}}, \bibinfo {author} {\bibfnamefont {J.~I.~A.}\ \bibnamefont {Li}}, \bibinfo {author} {\bibfnamefont {A.~A.}\ \bibnamefont {Zibrov}}, \bibinfo {author} {\bibfnamefont {L.}~\bibnamefont {Wang}}, \bibinfo {author} {\bibfnamefont {T.}~\bibnamefont {Taniguchi}}, \bibinfo {author} {\bibfnamefont {K.}~\bibnamefont {Watanabe}}, \bibinfo {author} {\bibfnamefont {J.~C.}\ \bibnamefont {Hone}}, \bibinfo {author} {\bibfnamefont {C.~R.}\ \bibnamefont {Dean}}, \bibinfo {author} {\bibfnamefont {M.~P.}\ \bibnamefont {Zaletel}}, \bibinfo {author} {\bibfnamefont {R.~C.}\ \bibnamefont {Ashoori}}, \bibinfo {author} {\bibfnamefont {A.~F.}\ \bibnamefont {Young}},\ and\ \bibinfo {author} {\bibfnamefont {A.~F.}\ \bibnamefont {Young}},\ }\bibfield  {title} {\bibinfo {title} {Direct measurement of discrete valley and orbital quantum numbers in bilayer graphene},\ }\href {https://www.nature.com/articles/s41467-017-00824-w} {\bibfield  {journal} {\bibinfo
  {journal} {Nature Communications}\ }\textbf {\bibinfo {volume} {8}} (\bibinfo {year} {2016})}\BibitemShut {NoStop}%
\bibitem [{\citenamefont {Wei}\ \emph {et~al.}(2025)\citenamefont {Wei}, \citenamefont {Xu}, \citenamefont {Villadiego},\ and\ \citenamefont {Huang}}]{Wei_Xu_Sodemann_Huang_LLM_SU4_breaking_MLG_2024}%
  \BibitemOpen
  \bibfield  {author} {\bibinfo {author} {\bibfnamefont {N.}~\bibnamefont {Wei}}, \bibinfo {author} {\bibfnamefont {G.}~\bibnamefont {Xu}}, \bibinfo {author} {\bibfnamefont {I.~S.}\ \bibnamefont {Villadiego}},\ and\ \bibinfo {author} {\bibfnamefont {C.}~\bibnamefont {Huang}},\ }\bibfield  {title} {\bibinfo {title} {Landau-level mixing and {S}{U}(4) symmetry breaking in graphene},\ }\href {https://doi.org/10.1103/PhysRevLett.134.046501} {\bibfield  {journal} {\bibinfo  {journal} {Phys. Rev. Lett.}\ }\textbf {\bibinfo {volume} {134}},\ \bibinfo {pages} {046501} (\bibinfo {year} {2025})}\BibitemShut {NoStop}%
\bibitem [{\citenamefont {Lian}\ and\ \citenamefont {Goerbig}(2017)}]{Lian_Goerbig_2017}%
  \BibitemOpen
  \bibfield  {author} {\bibinfo {author} {\bibfnamefont {Y.}~\bibnamefont {Lian}}\ and\ \bibinfo {author} {\bibfnamefont {M.~O.}\ \bibnamefont {Goerbig}},\ }\bibfield  {title} {\bibinfo {title} {Spin-valley skyrmions in graphene at filling factor $\ensuremath{\nu}=\ensuremath{-}1$},\ }\href {https://doi.org/10.1103/PhysRevB.95.245428} {\bibfield  {journal} {\bibinfo  {journal} {Phys. Rev. B}\ }\textbf {\bibinfo {volume} {95}},\ \bibinfo {pages} {245428} (\bibinfo {year} {2017})}\BibitemShut {NoStop}%
\bibitem [{\citenamefont {Atteia}\ and\ \citenamefont {Goerbig}(2021)}]{Atteia_Goerbig_2021}%
  \BibitemOpen
  \bibfield  {author} {\bibinfo {author} {\bibfnamefont {J.}~\bibnamefont {Atteia}}\ and\ \bibinfo {author} {\bibfnamefont {M.~O.}\ \bibnamefont {Goerbig}},\ }\bibfield  {title} {\bibinfo {title} {{S}{U}(4) spin waves in the $\ensuremath{\nu}=\ifmmode\pm\else\textpm\fi{}1$ quantum {Hall} ferromagnet in graphene},\ }\href {https://doi.org/10.1103/PhysRevB.103.195413} {\bibfield  {journal} {\bibinfo  {journal} {Phys. Rev. B}\ }\textbf {\bibinfo {volume} {103}},\ \bibinfo {pages} {195413} (\bibinfo {year} {2021})}\BibitemShut {NoStop}%
\bibitem [{\citenamefont {Farahi}\ \emph {et~al.}(2023)\citenamefont {Farahi}, \citenamefont {Chiu}, \citenamefont {Liu}, \citenamefont {Papic}, \citenamefont {Watanabe}, \citenamefont {Taniguchi}, \citenamefont {Zaletel},\ and\ \citenamefont {Yazdani}}]{Farahi23}%
  \BibitemOpen
  \bibfield  {author} {\bibinfo {author} {\bibfnamefont {G.}~\bibnamefont {Farahi}}, \bibinfo {author} {\bibfnamefont {C.-L.}\ \bibnamefont {Chiu}}, \bibinfo {author} {\bibfnamefont {X.}~\bibnamefont {Liu}}, \bibinfo {author} {\bibfnamefont {Z.}~\bibnamefont {Papic}}, \bibinfo {author} {\bibfnamefont {K.}~\bibnamefont {Watanabe}}, \bibinfo {author} {\bibfnamefont {T.}~\bibnamefont {Taniguchi}}, \bibinfo {author} {\bibfnamefont {M.~P.}\ \bibnamefont {Zaletel}},\ and\ \bibinfo {author} {\bibfnamefont {A.}~\bibnamefont {Yazdani}},\ }\bibfield  {title} {\bibinfo {title} {Broken symmetries and excitation spectra of interacting electrons in partially filled {Landau} levels},\ }\bibfield  {journal} {\bibinfo  {journal} {Nature Physics}\ }\href {https://doi.org/10.1038/s41567-023-02126-z} {10.1038/s41567-023-02126-z} (\bibinfo {year} {2023})\BibitemShut {NoStop}%
\bibitem [{\citenamefont {Pu}\ \emph {et~al.}(2022)\citenamefont {Pu}, \citenamefont {Balram},\ and\ \citenamefont {Papi\ifmmode~\acute{c}\else \'{c}\fi{}}}]{Pu22}%
  \BibitemOpen
  \bibfield  {author} {\bibinfo {author} {\bibfnamefont {S.}~\bibnamefont {Pu}}, \bibinfo {author} {\bibfnamefont {A.~C.}\ \bibnamefont {Balram}},\ and\ \bibinfo {author} {\bibfnamefont {Z.}~\bibnamefont {Papi\ifmmode~\acute{c}\else \'{c}\fi{}}},\ }\bibfield  {title} {\bibinfo {title} {Local density of states and particle entanglement in topological quantum fluids},\ }\href {https://doi.org/10.1103/PhysRevB.106.045140} {\bibfield  {journal} {\bibinfo  {journal} {Phys. Rev. B}\ }\textbf {\bibinfo {volume} {106}},\ \bibinfo {pages} {045140} (\bibinfo {year} {2022})}\BibitemShut {NoStop}%
\bibitem [{\citenamefont {Pu}\ \emph {et~al.}(2024)\citenamefont {Pu}, \citenamefont {Balram}, \citenamefont {Hu}, \citenamefont {Tsui}, \citenamefont {He}, \citenamefont {Regnault}, \citenamefont {Zaletel}, \citenamefont {Yazdani},\ and\ \citenamefont {Papi\ifmmode~\acute{c}\else \'{c}\fi{}}}]{Pu23a}%
  \BibitemOpen
  \bibfield  {author} {\bibinfo {author} {\bibfnamefont {S.}~\bibnamefont {Pu}}, \bibinfo {author} {\bibfnamefont {A.~C.}\ \bibnamefont {Balram}}, \bibinfo {author} {\bibfnamefont {Y.}~\bibnamefont {Hu}}, \bibinfo {author} {\bibfnamefont {Y.-C.}\ \bibnamefont {Tsui}}, \bibinfo {author} {\bibfnamefont {M.}~\bibnamefont {He}}, \bibinfo {author} {\bibfnamefont {N.}~\bibnamefont {Regnault}}, \bibinfo {author} {\bibfnamefont {M.~P.}\ \bibnamefont {Zaletel}}, \bibinfo {author} {\bibfnamefont {A.}~\bibnamefont {Yazdani}},\ and\ \bibinfo {author} {\bibfnamefont {Z.}~\bibnamefont {Papi\ifmmode~\acute{c}\else \'{c}\fi{}}},\ }\bibfield  {title} {\bibinfo {title} {Fingerprints of composite fermion {Lambda} levels in scanning tunneling microscopy},\ }\href {https://doi.org/10.1103/PhysRevB.110.L081107} {\bibfield  {journal} {\bibinfo  {journal} {Phys. Rev. B}\ }\textbf {\bibinfo {volume} {110}},\ \bibinfo {pages} {L081107} (\bibinfo {year} {2024})}\BibitemShut {NoStop}%
\bibitem [{\citenamefont {Gattu}\ \emph {et~al.}(2024)\citenamefont {Gattu}, \citenamefont {Sreejith},\ and\ \citenamefont {Jain}}]{Gattu23}%
  \BibitemOpen
  \bibfield  {author} {\bibinfo {author} {\bibfnamefont {M.}~\bibnamefont {Gattu}}, \bibinfo {author} {\bibfnamefont {G.~J.}\ \bibnamefont {Sreejith}},\ and\ \bibinfo {author} {\bibfnamefont {J.~K.}\ \bibnamefont {Jain}},\ }\bibfield  {title} {\bibinfo {title} {Scanning tunneling microscopy of fractional quantum {Hall} states: Spectroscopy of composite-fermion bound states},\ }\href {https://doi.org/10.1103/PhysRevB.109.L201123} {\bibfield  {journal} {\bibinfo  {journal} {Phys. Rev. B}\ }\textbf {\bibinfo {volume} {109}},\ \bibinfo {pages} {L201123} (\bibinfo {year} {2024})}\BibitemShut {NoStop}%
\bibitem [{\citenamefont {Kumar}\ \emph {et~al.}(2024{\natexlab{a}})\citenamefont {Kumar}, \citenamefont {Srivastav}, \citenamefont {Roy}, \citenamefont {Singhal}, \citenamefont {Watanabe}, \citenamefont {Taniguchi}, \citenamefont {Singh}, \citenamefont {Rolleau},\ and\ \citenamefont {Das}}]{Anindya_Das_2024_Heat_transport}%
  \BibitemOpen
  \bibfield  {author} {\bibinfo {author} {\bibfnamefont {R.}~\bibnamefont {Kumar}}, \bibinfo {author} {\bibfnamefont {S.~K.}\ \bibnamefont {Srivastav}}, \bibinfo {author} {\bibfnamefont {U.}~\bibnamefont {Roy}}, \bibinfo {author} {\bibfnamefont {U.}~\bibnamefont {Singhal}}, \bibinfo {author} {\bibfnamefont {K.}~\bibnamefont {Watanabe}}, \bibinfo {author} {\bibfnamefont {T.}~\bibnamefont {Taniguchi}}, \bibinfo {author} {\bibfnamefont {V.}~\bibnamefont {Singh}}, \bibinfo {author} {\bibfnamefont {P.}~\bibnamefont {Rolleau}},\ and\ \bibinfo {author} {\bibfnamefont {A.}~\bibnamefont {Das}},\ }\bibfield  {title} {\bibinfo {title} {Absence of heat flow in $\nu=0$ quantum hall ferromagnet in bilayer graphene},\ }\href {https://www.nature.com/articles/s41567-024-02673-z} {\bibfield  {journal} {\bibinfo  {journal} {Nature Physics}\ }\textbf {\bibinfo {volume} {20}},\ \bibinfo {pages} {1941} (\bibinfo {year} {2024}{\natexlab{a}})}\BibitemShut {NoStop}%
\bibitem [{\citenamefont {Delagrange}\ \emph {et~al.}(2024)\citenamefont {Delagrange}, \citenamefont {Garg}, \citenamefont {Le~Breton}, \citenamefont {Zhang}, \citenamefont {Dong}, \citenamefont {Jin}, \citenamefont {Watanabe}, \citenamefont {Taniguchi}, \citenamefont {Roulleau}, \citenamefont {Maillet}, \citenamefont {Roche},\ and\ \citenamefont {Parmentier}}]{Parmentier_2024_Heat_Flow}%
  \BibitemOpen
  \bibfield  {author} {\bibinfo {author} {\bibfnamefont {R.}~\bibnamefont {Delagrange}}, \bibinfo {author} {\bibfnamefont {M.}~\bibnamefont {Garg}}, \bibinfo {author} {\bibfnamefont {G.}~\bibnamefont {Le~Breton}}, \bibinfo {author} {\bibfnamefont {A.}~\bibnamefont {Zhang}}, \bibinfo {author} {\bibfnamefont {Q.}~\bibnamefont {Dong}}, \bibinfo {author} {\bibfnamefont {Y.}~\bibnamefont {Jin}}, \bibinfo {author} {\bibfnamefont {K.}~\bibnamefont {Watanabe}}, \bibinfo {author} {\bibfnamefont {T.}~\bibnamefont {Taniguchi}}, \bibinfo {author} {\bibfnamefont {P.}~\bibnamefont {Roulleau}}, \bibinfo {author} {\bibfnamefont {O.}~\bibnamefont {Maillet}}, \bibinfo {author} {\bibfnamefont {P.}~\bibnamefont {Roche}},\ and\ \bibinfo {author} {\bibfnamefont {F.~D.}\ \bibnamefont {Parmentier}},\ }\bibfield  {title} {\bibinfo {title} {Vanishing bulk heat flow in the $\nu$ = 0 quantum hall ferromagnet in monolayer graphene},\ }\href {https://doi.org/10.1038/s41567-024-02672-0} {\bibfield  {journal} {\bibinfo  {journal} {Nature
  Physics}\ }\textbf {\bibinfo {volume} {20}},\ \bibinfo {pages} {1927–1932} (\bibinfo {year} {2024})}\BibitemShut {NoStop}%
\bibitem [{\citenamefont {{Zibrov}}\ \emph {et~al.}(2017)\citenamefont {{Zibrov}}, \citenamefont {{Kometter}}, \citenamefont {{Zhou}}, \citenamefont {{Spanton}}, \citenamefont {{Taniguchi}}, \citenamefont {{Watanabe}}, \citenamefont {{Zaletel}},\ and\ \citenamefont {{Young}}}]{Zibrov16}%
  \BibitemOpen
  \bibfield  {author} {\bibinfo {author} {\bibfnamefont {A.~A.}\ \bibnamefont {{Zibrov}}}, \bibinfo {author} {\bibfnamefont {C.~R.}\ \bibnamefont {{Kometter}}}, \bibinfo {author} {\bibfnamefont {H.}~\bibnamefont {{Zhou}}}, \bibinfo {author} {\bibfnamefont {E.~M.}\ \bibnamefont {{Spanton}}}, \bibinfo {author} {\bibfnamefont {T.}~\bibnamefont {{Taniguchi}}}, \bibinfo {author} {\bibfnamefont {K.}~\bibnamefont {{Watanabe}}}, \bibinfo {author} {\bibfnamefont {M.~P.}\ \bibnamefont {{Zaletel}}},\ and\ \bibinfo {author} {\bibfnamefont {A.~F.}\ \bibnamefont {{Young}}},\ }\bibfield  {title} {\bibinfo {title} {Tunable interacting composite fermion phases in a half-filled bilayer-graphene {Landau} level},\ }\href {https://doi.org/10.1038/nature23893} {\bibfield  {journal} {\bibinfo  {journal} {Nature}\ }\textbf {\bibinfo {volume} {549}},\ \bibinfo {pages} {360} (\bibinfo {year} {2017})}\BibitemShut {NoStop}%
\bibitem [{\citenamefont {Li}\ \emph {et~al.}(2017)\citenamefont {Li}, \citenamefont {Tan}, \citenamefont {Chen}, \citenamefont {Zeng}, \citenamefont {Taniguchi}, \citenamefont {Watanabe}, \citenamefont {Hone},\ and\ \citenamefont {Dean}}]{Li17}%
  \BibitemOpen
  \bibfield  {author} {\bibinfo {author} {\bibfnamefont {J.~I.~A.}\ \bibnamefont {Li}}, \bibinfo {author} {\bibfnamefont {C.}~\bibnamefont {Tan}}, \bibinfo {author} {\bibfnamefont {S.}~\bibnamefont {Chen}}, \bibinfo {author} {\bibfnamefont {Y.}~\bibnamefont {Zeng}}, \bibinfo {author} {\bibfnamefont {T.}~\bibnamefont {Taniguchi}}, \bibinfo {author} {\bibfnamefont {K.}~\bibnamefont {Watanabe}}, \bibinfo {author} {\bibfnamefont {J.~C.}\ \bibnamefont {Hone}},\ and\ \bibinfo {author} {\bibfnamefont {C.~R.}\ \bibnamefont {Dean}},\ }\bibfield  {title} {\bibinfo {title} {Even-denominator fractional quantum {Hall} states in bilayer graphene},\ }\href {https://www.science.org/doi/10.1126/science.aao2521} {\bibfield  {journal} {\bibinfo  {journal} {Science}\ }\textbf {\bibinfo {volume} {358}},\ \bibinfo {pages} {648 } (\bibinfo {year} {2017})}\BibitemShut {NoStop}%
\bibitem [{\citenamefont {Assouline}\ \emph {et~al.}(2024)\citenamefont {Assouline}, \citenamefont {Wang}, \citenamefont {Zhou}, \citenamefont {Cohen}, \citenamefont {Yang}, \citenamefont {Zhang}, \citenamefont {Taniguchi}, \citenamefont {Watanabe}, \citenamefont {Mong}, \citenamefont {Zaletel},\ and\ \citenamefont {Young}}]{Assouline23}%
  \BibitemOpen
  \bibfield  {author} {\bibinfo {author} {\bibfnamefont {A.}~\bibnamefont {Assouline}}, \bibinfo {author} {\bibfnamefont {T.}~\bibnamefont {Wang}}, \bibinfo {author} {\bibfnamefont {H.}~\bibnamefont {Zhou}}, \bibinfo {author} {\bibfnamefont {L.~A.}\ \bibnamefont {Cohen}}, \bibinfo {author} {\bibfnamefont {F.}~\bibnamefont {Yang}}, \bibinfo {author} {\bibfnamefont {R.}~\bibnamefont {Zhang}}, \bibinfo {author} {\bibfnamefont {T.}~\bibnamefont {Taniguchi}}, \bibinfo {author} {\bibfnamefont {K.}~\bibnamefont {Watanabe}}, \bibinfo {author} {\bibfnamefont {R.~S.~K.}\ \bibnamefont {Mong}}, \bibinfo {author} {\bibfnamefont {M.~P.}\ \bibnamefont {Zaletel}},\ and\ \bibinfo {author} {\bibfnamefont {A.~F.}\ \bibnamefont {Young}},\ }\bibfield  {title} {\bibinfo {title} {Energy gap of the even-denominator fractional quantum {Hall} state in bilayer graphene},\ }\href {https://doi.org/10.1103/PhysRevLett.132.046603} {\bibfield  {journal} {\bibinfo  {journal} {Phys. Rev. Lett.}\ }\textbf {\bibinfo {volume} {132}},\ \bibinfo
  {pages} {046603} (\bibinfo {year} {2024})}\BibitemShut {NoStop}%
\bibitem [{\citenamefont {Kumar}\ \emph {et~al.}(2024{\natexlab{b}})\citenamefont {Kumar}, \citenamefont {Haug}, \citenamefont {Kim}, \citenamefont {Yutushui}, \citenamefont {Khudiakov}, \citenamefont {Bhardwaj}, \citenamefont {Ilin}, \citenamefont {Watanabe}, \citenamefont {Taniguchi}, \citenamefont {Mross},\ and\ \citenamefont {Ronen}}]{Kumar24}%
  \BibitemOpen
  \bibfield  {author} {\bibinfo {author} {\bibfnamefont {R.}~\bibnamefont {Kumar}}, \bibinfo {author} {\bibfnamefont {A.}~\bibnamefont {Haug}}, \bibinfo {author} {\bibfnamefont {J.}~\bibnamefont {Kim}}, \bibinfo {author} {\bibfnamefont {M.}~\bibnamefont {Yutushui}}, \bibinfo {author} {\bibfnamefont {K.}~\bibnamefont {Khudiakov}}, \bibinfo {author} {\bibfnamefont {V.}~\bibnamefont {Bhardwaj}}, \bibinfo {author} {\bibfnamefont {A.}~\bibnamefont {Ilin}}, \bibinfo {author} {\bibfnamefont {K.}~\bibnamefont {Watanabe}}, \bibinfo {author} {\bibfnamefont {T.}~\bibnamefont {Taniguchi}}, \bibinfo {author} {\bibfnamefont {D.~F.}\ \bibnamefont {Mross}},\ and\ \bibinfo {author} {\bibfnamefont {Y.}~\bibnamefont {Ronen}},\ }\href@noop {} {\bibinfo {title} {Quarter- and half-filled quantum {Hall} states and their competing interactions in bilayer graphene}} (\bibinfo {year} {2024}{\natexlab{b}}),\ \Eprint {https://arxiv.org/abs/2405.19405} {arXiv:2405.19405 [cond-mat.mes-hall]} \BibitemShut {NoStop}%
\bibitem [{\citenamefont {Chen}\ \emph {et~al.}(2024)\citenamefont {Chen}, \citenamefont {Huang}, \citenamefont {Li}, \citenamefont {Tong}, \citenamefont {Kuang}, \citenamefont {Xi}, \citenamefont {Watanabe}, \citenamefont {Taniguchi}, \citenamefont {Liu}, \citenamefont {Zhu}, \citenamefont {Lu}, \citenamefont {Zhang}, \citenamefont {Wu},\ and\ \citenamefont {Wang}}]{Chen24}%
  \BibitemOpen
  \bibfield  {author} {\bibinfo {author} {\bibfnamefont {Y.}~\bibnamefont {Chen}}, \bibinfo {author} {\bibfnamefont {Y.}~\bibnamefont {Huang}}, \bibinfo {author} {\bibfnamefont {Q.}~\bibnamefont {Li}}, \bibinfo {author} {\bibfnamefont {B.}~\bibnamefont {Tong}}, \bibinfo {author} {\bibfnamefont {G.}~\bibnamefont {Kuang}}, \bibinfo {author} {\bibfnamefont {C.}~\bibnamefont {Xi}}, \bibinfo {author} {\bibfnamefont {K.}~\bibnamefont {Watanabe}}, \bibinfo {author} {\bibfnamefont {T.}~\bibnamefont {Taniguchi}}, \bibinfo {author} {\bibfnamefont {G.}~\bibnamefont {Liu}}, \bibinfo {author} {\bibfnamefont {Z.}~\bibnamefont {Zhu}}, \bibinfo {author} {\bibfnamefont {L.}~\bibnamefont {Lu}}, \bibinfo {author} {\bibfnamefont {F.-C.}\ \bibnamefont {Zhang}}, \bibinfo {author} {\bibfnamefont {Y.-H.}\ \bibnamefont {Wu}},\ and\ \bibinfo {author} {\bibfnamefont {L.}~\bibnamefont {Wang}},\ }\bibfield  {title} {\bibinfo {title} {Tunable even- and odd-denominator fractional quantum {Hall} states in trilayer graphene},\ }\href
  {https://doi.org/10.1038/s41467-024-50589-2} {\bibfield  {journal} {\bibinfo  {journal} {Nature Communications}\ }\textbf {\bibinfo {volume} {15}},\ \bibinfo {pages} {6236} (\bibinfo {year} {2024})}\BibitemShut {NoStop}%
\bibitem [{\citenamefont {Chanda}\ \emph {et~al.}(2025)\citenamefont {Chanda}, \citenamefont {Kaur}, \citenamefont {Singh}, \citenamefont {Watanabe}, \citenamefont {Taniguchi}, \citenamefont {Jain}, \citenamefont {Khanna}, \citenamefont {Balram},\ and\ \citenamefont {Bid}}]{Chanda25}%
  \BibitemOpen
  \bibfield  {author} {\bibinfo {author} {\bibfnamefont {T.}~\bibnamefont {Chanda}}, \bibinfo {author} {\bibfnamefont {S.}~\bibnamefont {Kaur}}, \bibinfo {author} {\bibfnamefont {H.}~\bibnamefont {Singh}}, \bibinfo {author} {\bibfnamefont {K.}~\bibnamefont {Watanabe}}, \bibinfo {author} {\bibfnamefont {T.}~\bibnamefont {Taniguchi}}, \bibinfo {author} {\bibfnamefont {M.}~\bibnamefont {Jain}}, \bibinfo {author} {\bibfnamefont {U.}~\bibnamefont {Khanna}}, \bibinfo {author} {\bibfnamefont {A.~C.}\ \bibnamefont {Balram}},\ and\ \bibinfo {author} {\bibfnamefont {A.}~\bibnamefont {Bid}},\ }\href {https://arxiv.org/abs/2502.06245} {\bibinfo {title} {Even denominator fractional quantum {Hall} states in the zeroth {Landau} level of monolayer-like band of {A}{B}{A} trilayer graphene}} (\bibinfo {year} {2025}),\ \Eprint {https://arxiv.org/abs/2502.06245} {arXiv:2502.06245 [cond-mat.mes-hall]} \BibitemShut {NoStop}%
\bibitem [{\citenamefont {Cai}\ \emph {et~al.}(2023)\citenamefont {Cai}, \citenamefont {Anderson}, \citenamefont {Wang}, \citenamefont {Zhang}, \citenamefont {Liu}, \citenamefont {Holtzmann}, \citenamefont {Zhang}, \citenamefont {Fan}, \citenamefont {Taniguchi}, \citenamefont {Watanabe}, \citenamefont {Ran}, \citenamefont {Cao}, \citenamefont {Fu}, \citenamefont {Xiao}, \citenamefont {Yao},\ and\ \citenamefont {Xu}}]{FQAH_MoTe2_Xu_2023a}%
  \BibitemOpen
  \bibfield  {author} {\bibinfo {author} {\bibfnamefont {J.}~\bibnamefont {Cai}}, \bibinfo {author} {\bibfnamefont {E.}~\bibnamefont {Anderson}}, \bibinfo {author} {\bibfnamefont {C.}~\bibnamefont {Wang}}, \bibinfo {author} {\bibfnamefont {X.}~\bibnamefont {Zhang}}, \bibinfo {author} {\bibfnamefont {X.}~\bibnamefont {Liu}}, \bibinfo {author} {\bibfnamefont {W.}~\bibnamefont {Holtzmann}}, \bibinfo {author} {\bibfnamefont {Y.}~\bibnamefont {Zhang}}, \bibinfo {author} {\bibfnamefont {F.}~\bibnamefont {Fan}}, \bibinfo {author} {\bibfnamefont {T.}~\bibnamefont {Taniguchi}}, \bibinfo {author} {\bibfnamefont {K.}~\bibnamefont {Watanabe}}, \bibinfo {author} {\bibfnamefont {Y.}~\bibnamefont {Ran}}, \bibinfo {author} {\bibfnamefont {T.}~\bibnamefont {Cao}}, \bibinfo {author} {\bibfnamefont {L.}~\bibnamefont {Fu}}, \bibinfo {author} {\bibfnamefont {D.}~\bibnamefont {Xiao}}, \bibinfo {author} {\bibfnamefont {W.}~\bibnamefont {Yao}},\ and\ \bibinfo {author} {\bibfnamefont {X.}~\bibnamefont {Xu}},\ }\bibfield  {title}
  {\bibinfo {title} {Signatures of fractional quantum anomalous {Hall} states in twisted {M}o{T}e2},\ }\href {https://doi.org/10.1038/s41586-023-06289-w} {\bibfield  {journal} {\bibinfo  {journal} {Nature}\ }\textbf {\bibinfo {volume} {622}},\ \bibinfo {pages} {63} (\bibinfo {year} {2023})}\BibitemShut {NoStop}%
\bibitem [{\citenamefont {Park}\ \emph {et~al.}(2023)\citenamefont {Park}, \citenamefont {Cai}, \citenamefont {Anderson}, \citenamefont {Zhang}, \citenamefont {Zhu}, \citenamefont {Liu}, \citenamefont {Wang}, \citenamefont {Holtzmann}, \citenamefont {Hu}, \citenamefont {Liu}, \citenamefont {Taniguchi}, \citenamefont {Watanabe}, \citenamefont {Chu}, \citenamefont {Cao}, \citenamefont {Fu}, \citenamefont {Yao}, \citenamefont {Chang}, \citenamefont {Cobden}, \citenamefont {Xiao},\ and\ \citenamefont {Xu}}]{FQAH_MoTe2_Xu_2023b}%
  \BibitemOpen
  \bibfield  {author} {\bibinfo {author} {\bibfnamefont {H.}~\bibnamefont {Park}}, \bibinfo {author} {\bibfnamefont {J.}~\bibnamefont {Cai}}, \bibinfo {author} {\bibfnamefont {E.}~\bibnamefont {Anderson}}, \bibinfo {author} {\bibfnamefont {Y.}~\bibnamefont {Zhang}}, \bibinfo {author} {\bibfnamefont {J.}~\bibnamefont {Zhu}}, \bibinfo {author} {\bibfnamefont {X.}~\bibnamefont {Liu}}, \bibinfo {author} {\bibfnamefont {C.}~\bibnamefont {Wang}}, \bibinfo {author} {\bibfnamefont {W.}~\bibnamefont {Holtzmann}}, \bibinfo {author} {\bibfnamefont {C.}~\bibnamefont {Hu}}, \bibinfo {author} {\bibfnamefont {Z.}~\bibnamefont {Liu}}, \bibinfo {author} {\bibfnamefont {T.}~\bibnamefont {Taniguchi}}, \bibinfo {author} {\bibfnamefont {K.}~\bibnamefont {Watanabe}}, \bibinfo {author} {\bibfnamefont {J.-H.}\ \bibnamefont {Chu}}, \bibinfo {author} {\bibfnamefont {T.}~\bibnamefont {Cao}}, \bibinfo {author} {\bibfnamefont {L.}~\bibnamefont {Fu}}, \bibinfo {author} {\bibfnamefont {W.}~\bibnamefont {Yao}}, \bibinfo {author}
  {\bibfnamefont {C.-Z.}\ \bibnamefont {Chang}}, \bibinfo {author} {\bibfnamefont {D.}~\bibnamefont {Cobden}}, \bibinfo {author} {\bibfnamefont {D.}~\bibnamefont {Xiao}},\ and\ \bibinfo {author} {\bibfnamefont {X.}~\bibnamefont {Xu}},\ }\bibfield  {title} {\bibinfo {title} {Observation of fractionally quantized anomalous {Hall} effect},\ }\href {https://doi.org/10.1038/s41586-023-06536-0} {\bibfield  {journal} {\bibinfo  {journal} {Nature}\ }\textbf {\bibinfo {volume} {622}},\ \bibinfo {pages} {74} (\bibinfo {year} {2023})}\BibitemShut {NoStop}%
\bibitem [{\citenamefont {Zeng}\ \emph {et~al.}(2023)\citenamefont {Zeng}, \citenamefont {Xia}, \citenamefont {Kang}, \citenamefont {Zhu}, \citenamefont {Kn{\"u}ppel}, \citenamefont {Vaswani}, \citenamefont {Watanabe}, \citenamefont {Taniguchi}, \citenamefont {Mak},\ and\ \citenamefont {Shan}}]{FQAH_MoTe2_Mak_Shan_2023}%
  \BibitemOpen
  \bibfield  {author} {\bibinfo {author} {\bibfnamefont {Y.}~\bibnamefont {Zeng}}, \bibinfo {author} {\bibfnamefont {Z.}~\bibnamefont {Xia}}, \bibinfo {author} {\bibfnamefont {K.}~\bibnamefont {Kang}}, \bibinfo {author} {\bibfnamefont {J.}~\bibnamefont {Zhu}}, \bibinfo {author} {\bibfnamefont {P.}~\bibnamefont {Kn{\"u}ppel}}, \bibinfo {author} {\bibfnamefont {C.}~\bibnamefont {Vaswani}}, \bibinfo {author} {\bibfnamefont {K.}~\bibnamefont {Watanabe}}, \bibinfo {author} {\bibfnamefont {T.}~\bibnamefont {Taniguchi}}, \bibinfo {author} {\bibfnamefont {K.~F.}\ \bibnamefont {Mak}},\ and\ \bibinfo {author} {\bibfnamefont {J.}~\bibnamefont {Shan}},\ }\bibfield  {title} {\bibinfo {title} {Thermodynamic evidence of fractional {Chern} insulator in moir{\'e} {M}o{T}e2},\ }\href {https://doi.org/10.1038/s41586-023-06452-3} {\bibfield  {journal} {\bibinfo  {journal} {Nature}\ }\textbf {\bibinfo {volume} {622}},\ \bibinfo {pages} {69} (\bibinfo {year} {2023})}\BibitemShut {NoStop}%
\bibitem [{\citenamefont {Xu}\ \emph {et~al.}(2023)\citenamefont {Xu}, \citenamefont {Sun}, \citenamefont {Jia}, \citenamefont {Liu}, \citenamefont {Xu}, \citenamefont {Li}, \citenamefont {Gu}, \citenamefont {Watanabe}, \citenamefont {Taniguchi}, \citenamefont {Tong}, \citenamefont {Jia}, \citenamefont {Shi}, \citenamefont {Jiang}, \citenamefont {Zhang}, \citenamefont {Liu},\ and\ \citenamefont {Li}}]{FQAHE_MoTe2_Li_2023}%
  \BibitemOpen
  \bibfield  {author} {\bibinfo {author} {\bibfnamefont {F.}~\bibnamefont {Xu}}, \bibinfo {author} {\bibfnamefont {Z.}~\bibnamefont {Sun}}, \bibinfo {author} {\bibfnamefont {T.}~\bibnamefont {Jia}}, \bibinfo {author} {\bibfnamefont {C.}~\bibnamefont {Liu}}, \bibinfo {author} {\bibfnamefont {C.}~\bibnamefont {Xu}}, \bibinfo {author} {\bibfnamefont {C.}~\bibnamefont {Li}}, \bibinfo {author} {\bibfnamefont {Y.}~\bibnamefont {Gu}}, \bibinfo {author} {\bibfnamefont {K.}~\bibnamefont {Watanabe}}, \bibinfo {author} {\bibfnamefont {T.}~\bibnamefont {Taniguchi}}, \bibinfo {author} {\bibfnamefont {B.}~\bibnamefont {Tong}}, \bibinfo {author} {\bibfnamefont {J.}~\bibnamefont {Jia}}, \bibinfo {author} {\bibfnamefont {Z.}~\bibnamefont {Shi}}, \bibinfo {author} {\bibfnamefont {S.}~\bibnamefont {Jiang}}, \bibinfo {author} {\bibfnamefont {Y.}~\bibnamefont {Zhang}}, \bibinfo {author} {\bibfnamefont {X.}~\bibnamefont {Liu}},\ and\ \bibinfo {author} {\bibfnamefont {T.}~\bibnamefont {Li}},\ }\bibfield  {title} {\bibinfo {title}
  {Observation of integer and fractional quantum anomalous {Hall} effects in twisted bilayer ${\mathrm{{m}o{t}e}}_{2}$},\ }\href {https://doi.org/10.1103/PhysRevX.13.031037} {\bibfield  {journal} {\bibinfo  {journal} {Phys. Rev. X}\ }\textbf {\bibinfo {volume} {13}},\ \bibinfo {pages} {031037} (\bibinfo {year} {2023})}\BibitemShut {NoStop}%
\bibitem [{\citenamefont {Lu}\ \emph {et~al.}(2024)\citenamefont {Lu}, \citenamefont {Han}, \citenamefont {Yao}, \citenamefont {Reddy}, \citenamefont {Yang}, \citenamefont {Seo}, \citenamefont {Watanabe}, \citenamefont {Taniguchi}, \citenamefont {Fu},\ and\ \citenamefont {Ju}}]{FQAH_Pentalayer_Graphene_Ju_2024}%
  \BibitemOpen
  \bibfield  {author} {\bibinfo {author} {\bibfnamefont {Z.}~\bibnamefont {Lu}}, \bibinfo {author} {\bibfnamefont {T.}~\bibnamefont {Han}}, \bibinfo {author} {\bibfnamefont {Y.}~\bibnamefont {Yao}}, \bibinfo {author} {\bibfnamefont {A.~P.}\ \bibnamefont {Reddy}}, \bibinfo {author} {\bibfnamefont {J.}~\bibnamefont {Yang}}, \bibinfo {author} {\bibfnamefont {J.}~\bibnamefont {Seo}}, \bibinfo {author} {\bibfnamefont {K.}~\bibnamefont {Watanabe}}, \bibinfo {author} {\bibfnamefont {T.}~\bibnamefont {Taniguchi}}, \bibinfo {author} {\bibfnamefont {L.}~\bibnamefont {Fu}},\ and\ \bibinfo {author} {\bibfnamefont {L.}~\bibnamefont {Ju}},\ }\bibfield  {title} {\bibinfo {title} {Fractional quantum anomalous {Hall} effect in multilayer graphene},\ }\href {https://doi.org/10.1038/s41586-023-07010-7} {\bibfield  {journal} {\bibinfo  {journal} {Nature}\ }\textbf {\bibinfo {volume} {626}},\ \bibinfo {pages} {759} (\bibinfo {year} {2024})}\BibitemShut {NoStop}%
\bibitem [{dia()}]{diagham}%
  \BibitemOpen
  \href@noop {} {}\bibinfo {note} {Diag{H}am, \url{https://www.nick-ux.org/diagham}}\BibitemShut {NoStop}%
\end{thebibliography}%

\clearpage


\section*{Supplementary material}

# Supplemental Material for "Anomalous Transport Gaps of Fractional Quantum Hall Phases in Graphene Landau Levels are Induced by Spin-Valley Entangled Ground States"

Jincheng An,[1, *] Ajit C. Balram,[2, 3, †] Udit Khanna,[4, ‡] and Ganpathy Murthy[1, §]

[1]*Department of Physics and Astronomy, University of Kentucky, Lexington, KY 40506, USA*
[2]*Institute of Mathematical Sciences, CIT Campus, Chennai, 600113, India*
[3]*Homi Bhabha National Institute, Training School Complex, Anushaktinagar, Mumbai 400094, India*
[4]*Theoretical Physics Division, Physical Research Laboratory, Navrangpura, Ahmedabad-380009, India*

This supplemental material presents the technical details of our analysis used to obtain the results presented in the main text. In Sec. SI, we present the variational energy functional for fractional quantum Hall (FQH) states at filling $(1, \nu_F)$ in the **n**=0 and **n**=1 Landau levels (LL) of graphene. We also introduce the spinor ansatz used for energy minimization, as well as the order parameters that characterize all the possible phases. In Sec. SII, we present the spinors for all phases that appear in our analysis at filling $(1, \nu_F)$. In Sec. SIII, we present the formalism for evaluating the transport gaps of FQH states at filling $(1, \nu_F)$. Finally, in Sec. SIV we present reasons why the most obvious candidates for producing the anomalous $E_Z$ dependence in the 1LLs [1], namely, the VPAFM or V/SAFM phases, cannot possibly do so.

## SI. VARIATIONAL ENERGY FUNCTIONAL, SPINOR ANSATZ, ORDER PARAMETERS

The $B_\perp = 0$ Hamiltonian describing ultra-short-range (USR) anisotropic interactions may be written as [2],

$$\hat{H}^{\mathrm{an}} = \frac{1}{2} \sum_{a,b} g_{ab} \int d^2\mathbf{r} : \left[\hat{\psi}^\dagger(\mathbf{r})\left(\tau^a_{\mathrm{valley}} \otimes \eta^b_{\mathrm{sublattice}}\right)\hat{\psi}(\mathbf{r})\right]^2 :, \tag{S1}$$

where $\hat{\psi}$ is an 8-component local field operator (combining spin, valley and sublattice degrees of freedom), $\tau(\eta)$ are Pauli matrices defined in valley (sublattice) space with $a$, $b = 0$, $x$, $y$, $z$. The couplings $g_{ab}$'s satisfy the following constraints, leading to 8 independent parameters ($g_{00}$ is absorbed in the Coulomb interaction),

$$\begin{aligned} g_{x0} &= g_{y0} \equiv g_{\perp 0}, \ g_{xz} = g_{yz} \equiv g_{\perp z}, \\ g_{0x} &= g_{0y} \equiv g_{0\perp}, \ g_{zx} = g_{zy} \equiv g_{z\perp}, \\ g_{xx} &= g_{xy} = g_{yx} = g_{yy} \equiv g_{\perp\perp}. \end{aligned} \tag{S2}$$

In the symmetric gauge (at $B_\perp \neq 0$), the Hamiltonian projected to the **n**-th LL of graphene is [3],

$$\hat{H}^{\mathrm{an}}_{\mathbf{n}} = \frac{1}{2} \sum_{a=x,y,z} \sum_m u^{(m)}_{a,\mathbf{n}} \mathbf{U}^{(m)}_{m_1 m_2 m_3 m_4} : \hat{\tau}^a_{m_1 m_4} \hat{\tau}^a_{m_2 m_3} :, \tag{S3}$$

$$\hat{\tau}^a_{m_1 m_4} = (\tau^a \otimes \sigma^0_{\mathrm{spin}})_{\alpha\beta} \hat{c}^\dagger_{m_1,\alpha} \hat{c}_{m_4,\beta}, \quad \alpha, \beta = K\uparrow,\ K\downarrow,\ K'\uparrow,\ K'\downarrow,$$

where $u^{(m)}_{x,\mathbf{n}} = u^{(m)}_{x,\mathbf{n}} \equiv u^{(m)}_{\perp,\mathbf{n}}$ and $u^{(m)}_{z,\mathbf{n}}$ are the Haldane pseudopotentials [4] in the relative angular momentum channel $m$, in the **n**-th LL, and $\mathbf{U}^{(m)}_{m_1 m_2 m_3 m_4}$'s are the corresponding interaction matrix elements which satisfy [3],

$$\mathbf{U}^{(m)}_{m_1 m_2 m_2 m_1} = (-1)^m \mathbf{U}^{(m)}_{m_1 m_2 m_1 m_2}, \quad \frac{1}{N_\phi} \sum_{m_1 m_2} \mathbf{U}^{(m)}_{m_1 m_2 m_2 m_1} = 2. \tag{S4}$$

---

[*] jincheng.an1@gmail.com
[†] cb.ajit@gmail.com
[‡] udit.khanna.10@gmail.com
[§] murthy@g.uky.edu

In the **n**-th LL of graphene, USR interactions generate a finite value for the $m \leq 2n$ pseudopotentials. In **n** = 0 and 1 LLs, these are related to the microscopic couplings ($g_{ab}$) through [5],

$$u^{(0)}_{\perp,\mathbf{n}=0} = \frac{g_{\perp 0} + g_{\perp z}}{4\pi\ell^2} \equiv \frac{g_\perp}{4\pi\ell^2}, \quad u^{(0)}_{z,\mathbf{n}=0} = \frac{g_{z0} + g_{zz}}{4\pi\ell^2} \equiv \frac{g_z}{4\pi\ell^2}, \tag{S5}$$

$$\begin{cases} u^{(0)}_{\perp,\mathbf{n}=1} = \frac{5g_{\perp 0} + g_{\perp z} + 4g_{\perp\perp}}{32\pi\ell^2}, \\ u^{(1)}_{\perp,\mathbf{n}=1} = \frac{g_{\perp 0} - g_{\perp z} - 2g_{\perp\perp}}{16\pi\ell^2}, \\ u^{(2)}_{\perp,\mathbf{n}=1} = \frac{g_{\perp 0} + g_{\perp z}}{32\pi\ell^2}, \end{cases} \quad \begin{cases} u^{(0)}_{z,\mathbf{n}=1} = \frac{5g_{z0} + g_{zz} + 4g_{z\perp}}{32\pi\ell^2}, \\ u^{(1)}_{z,\mathbf{n}=1} = \frac{g_{z0} - g_{zz} - 2g_{z\perp}}{16\pi\ell^2}, \\ u^{(2)}_{z,\mathbf{n}=1} = \frac{g_{z0} + g_{zz}}{32\pi\ell^2}. \end{cases} \tag{S6}$$

In the **n** = 0 LL, these interactions are controlled by just two independent couplings: $g_\perp = g_{\perp 0} + g_{\perp z}$, $g_z = g_{z0} + g_{zz}$, whereas in the **n** = 1 LL, all six independent couplings enter the interaction Hamiltonian.

To study QHFM within any manifold of four degenerate LLs, we describe the phase at filling $n + \nu_F$ (where $n = 0, 1, 2, 3$; $0 < \nu_F < 1$) with variational states of the form $|\Psi\rangle = |I\rangle \otimes |F\rangle$ where $|I\rangle$ denotes a Slater determinant state of $n$ fully occupied LLs and $|F\rangle$ is a FQH state with filling $\nu_F$. The spin-valley spinors of fully occupied levels are denoted by $|f_1\rangle, ..., |f_n\rangle$, while $|f_{n+1}\rangle$ denotes the spinor of the fractionally filled level. The directions of these spinors are the variational parameters in our approach, i.e., the ground state corresponds to the set of spinors entering $|\Psi\rangle$ that minimizes the energy with respect to the anisotropic interactions in Eq. (S3). Following the procedure in Refs. [3, 5], one may write the variational energy functional as,

$$\frac{\langle\Psi|\hat{H}^{\text{an}}_{\mathbf{n}}|\Psi\rangle}{N_\phi} = E_{II}(P_I, P_I) + E_{IF}(P_I, P_F) + E_{FF}(P_F, P_F), \tag{S7}$$

$$E_{II} = \frac{1}{2}\sum_m 2\mathcal{E}^{\text{an}}_{HF}(P_I, P_I; \mathbf{n}, m), \quad E_{IF} = \frac{1}{2} \times 2\nu_F \sum_m 2\mathcal{E}^{\text{an}}_{HF}(P_I, P_F; \mathbf{n}, m), \tag{S8}$$

$$E_{FF} = \frac{1}{2}\sum_{\text{odd } m} \langle \mathbf{U}^{(m)} \rangle_{\nu_F} \times \frac{1}{2}\mathcal{E}^{\text{an}}_{HF}(P_F, P_F; \mathbf{n}, m), \tag{S9}$$

where $P_I = \sum_{i=1}^n |f_i\rangle\langle f_i|$ and $P_F = |f_{n+1}\rangle\langle f_{n+1}|$ are projectors into the subspace of fully filled and fractionally filled states. $\langle \hat{\mathbf{U}}^{(m)} \rangle_{\nu_F} \equiv \langle \nu_F | \mathbf{U}^m_{m_1 m_2 m_3 m_4} \hat{c}^\dagger_{m_1} \hat{c}^\dagger_{m_2} \hat{c}_{m_3} \hat{c}_{m_4} | \nu_F \rangle / N_\phi$ is the average pair amplitude with relative angular momentum $m$ in the FQH state $|\nu_F\rangle$, and $\mathcal{E}^{\text{an}}_{HF}$ is the Hartree-Fock energy functional for an interaction with Haldane pseudopotential $V_m$ in the **n**-th LL,

$$\mathcal{E}^{\text{an}}_{HF}(P_1, P_2; \mathbf{n}, m) = \sum_{a=x,y,z} u^{(m)}_{a,\mathbf{n}} \Big[\text{Tr}(\tau^a P_1)\text{Tr}(\tau^a P_2) - (-1)^m \text{Tr}(\tau^a P_1 \tau^a P_2)\Big]. \tag{S10}$$

The possible phases at a given filling factor may be obtained by minimizing the energy functional over the space of couplings ($g$'s) using the following spinor ansatz,

$$|f_1\rangle = \cos\frac{\alpha_1}{2}|\boldsymbol{\tau}\rangle\otimes|\mathbf{s}_a\rangle + \sin\frac{\alpha_1}{2}|-\boldsymbol{\tau}\rangle\otimes|-\mathbf{s}_b\rangle, \quad |f_2\rangle = \cos\frac{\alpha_2}{2}|\boldsymbol{\tau}\rangle\otimes|-\mathbf{s}_a\rangle + \sin\frac{\alpha_2}{2}|-\boldsymbol{\tau}\rangle\otimes|\mathbf{s}_b\rangle,$$

$$|f_3\rangle = \sin\frac{\alpha_1}{2}|\boldsymbol{\tau}\rangle\otimes|\mathbf{s}_a\rangle - \cos\frac{\alpha_1}{2}|-\boldsymbol{\tau}\rangle\otimes|-\mathbf{s}_b\rangle, \quad |f_4\rangle = \sin\frac{\alpha_2}{2}|\boldsymbol{\tau}\rangle\otimes|-\mathbf{s}_a\rangle - \cos\frac{\alpha_2}{2}|-\boldsymbol{\tau}\rangle\otimes|\mathbf{s}_b\rangle,$$

$$|\boldsymbol{\tau}\rangle = \begin{pmatrix} \cos\frac{\theta_\tau}{2} \\ e^{i\phi_\tau}\sin\frac{\theta_\tau}{2} \end{pmatrix}, \qquad |\mathbf{s}\rangle = \begin{pmatrix} \cos\frac{\theta_s}{2} \\ e^{i\phi_s}\sin\frac{\theta_s}{2} \end{pmatrix}, \tag{S11}$$

with $|\boldsymbol{\tau}(\mathbf{s})\rangle$ being unit vectors on the valley (spin) Bloch spheres (see Refs. [3, 5, 6] for details). Since our analysis is restricted to the **n** = 0 & 1 LLs, we are only concerned with the $m \leq 2$ pair amplitudes. For the Laughlin state at $\nu_F = 1/3$, the pair amplitudes with $m < 3$ are zero, so that $\langle \mathbf{U}^{(m)} \rangle_{\frac{1}{3}} = 0$ for all $m$ that enter our analysis. For the one-component state at $\nu_F = 2/3$, the only nonzero pair amplitude is, $\langle \mathbf{U}^{(1)} \rangle_{\frac{2}{3}} = 4/3$ [5].

The different phases at any given filling $\vec{\nu} = (\nu_1, \nu_2, .., \nu_i, ...)$ [where, $\nu_i$ is the filling of spinor $|f_i\rangle$] may be characterized through expectation values of order parameters $\hat{O}$,

$$\langle \hat{O} \rangle = \frac{\sum_i \nu_i \text{Tr}(P_i \hat{O})}{\sum_i \nu_i}. \tag{S12}$$

Here, $P_i = |f_i\rangle\langle f_i|$ (not to be confused with $P_I$) is the projector onto the spinor $|f_i\rangle$. The order parameters relevant to our analysis are: $\sigma_z$ [representing ferromagnetic (FM) order], $\tau_z \sigma_x$ [representing canted antiferromagnetic (CAFM)

order], $\tau_z$ [representing valley polarized (VP) order, which can also be interpreted as charge-density-wave (CDW) order in the $\mathbf{n} = 0$ LL due to spin-valley locking], $\tau_x$ [representing intervalley coherence or bond order (BO), sometimes also termed Kekulé distorted (KD) order], $\tau_x\sigma_x$ [representing spin-valley entangled-X (SVEX) order], $\tau_y\sigma_y$ (SVEY), and $\tau_z\sigma_z$ [representing antiferromagnetic (AFM) order].

## SII. POSSIBLE PHASES OF FQH STATES AT $(1,\nu_F)$

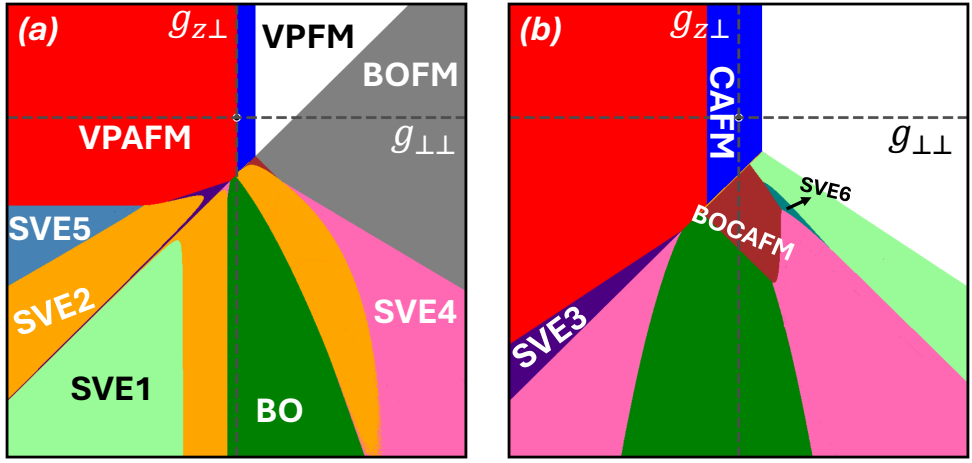

FIG. S1: Phase diagrams of filling (a) $(1,\frac{1}{3})$ and (b) $(1,\frac{2}{3})$ in $\mathbf{n} = 1$ LL with $g_{\perp\perp} \in [-1600, 1600]$meV·nm$^2$, $g_{z\perp} \in [-2400, 800]$meV·nm$^2$ at $B_{\text{tot}} = B_\perp = 17$T.

For the state with filling factor $(1,\nu_F)$ [$\nu_F = 1/3, 2/3$] we minimized the energy functional described in the previous section over a large region of the space of couplings in both the $\mathbf{n} = 0$ and 1 LLs. The optimal values of the angles parameterizing the spinors [Eq. S11] were determined analytically for the simpler phases and numerically for others. This analysis yielded a rich set of phase diagrams which are qualitatively different for the $\mathbf{n} = 0$ and 1 LLs, as well as for different $\nu$ (see [3, 5] for more details). Figure S1 presents two representative phase diagrams over the $(g_{\perp\perp}, g_{z\perp})$ space for fillings $(1,\frac{1}{3})$ and $(1,\frac{2}{3})$ in $\mathbf{n} = 1$ LL, where we have fixed the other 4 couplings $(g_{\perp z}, g_{zz}, g_{\perp 0}, g_{z0}) = (-600, 200, 0, 0)$ meV nm$^2$. The spinors in each phase are summarized below, with the first spinor being fully filled and the second partially filled:

**(1)** ○ Valley Polarized Ferromagnet(VPFM): $|K\rangle\otimes|\uparrow\rangle$, $|K'\rangle\otimes|\uparrow\rangle$

**(2)** ● Valley Polarized Antiferromagnet(VPAFM): $|K\rangle\otimes|\uparrow\rangle$, $|K\rangle\otimes|\downarrow\rangle$

**(3)** ○ Valley/Spin Antiferromagnet(V/SAFM): $|K\rangle\otimes|\uparrow\rangle$, $|K'\rangle\otimes|\downarrow\rangle$

**(4)** ● Canted Antiferromagnet(CAFM): $|K\rangle\otimes|\nwarrow_1\rangle$, $|K'\rangle\otimes|\nearrow_2\rangle$

**(5)** ● Bond-Ordered Ferromagnet(BOFM): $|\hat{e}_x\rangle\otimes|\uparrow\rangle$, $|-\hat{e}_x\rangle\otimes|\uparrow\rangle$

**(6)** ● Bond Order(BO): $|\hat{e}_x\rangle\otimes|\uparrow\rangle$, $|\hat{e}_x\rangle\otimes|\downarrow\rangle$

**(7)** ● Bond-ordered CAFM(BOCAFM): $|\hat{e}_x\rangle\otimes|\nwarrow_1\rangle$, $|-\hat{e}_x\rangle\otimes|\nearrow_2\rangle$

**(8)** ● Spin-valley entangled 1 (SVE1): $\cos\frac{\alpha}{2}|K'\rangle\otimes|\uparrow\rangle - \sin\frac{\alpha}{2}|K\rangle\otimes|\downarrow\rangle$, $|K\rangle\otimes|\uparrow\rangle$

**(9)** ● SVE2: No simple description for this phase, all order parameters are nonzero.

**(10)** ● SVE3: $\cos\frac{\alpha}{2}|K\rangle\otimes|\uparrow\rangle - \sin\frac{\alpha}{2}|K'\rangle\otimes|\downarrow\rangle$, $|K\rangle\otimes|\downarrow\rangle$

**(11)** 🟣 SVE4: $\cos\frac{\alpha_1}{2}|\hat{e}_x\rangle\otimes|\uparrow\rangle - \sin\frac{\alpha_1}{2}|-\hat{e}_x\rangle\otimes|\downarrow\rangle$, $\cos\frac{\alpha_2}{2}|\hat{e}_x\rangle\otimes|\downarrow\rangle - \sin\frac{\alpha_2}{2}|-\hat{e}_x\rangle\otimes|\uparrow\rangle$

**(12)** 🔵 SVE5: $\cos\frac{\alpha_1}{2}|K\rangle\otimes|\uparrow\rangle - \sin\frac{\alpha_1}{2}|K'\rangle\otimes|\downarrow\rangle$, $\sin\frac{\alpha_1}{2}|K\rangle\otimes|\uparrow\rangle + \cos\frac{\alpha_1}{2}|K'\rangle\otimes|\downarrow\rangle$

**(13)** 🟢 SVE6: $\cos\frac{\alpha_1}{2}|K\rangle\otimes|\downarrow\rangle - \sin\frac{\alpha_1}{2}|K'\rangle\otimes|\uparrow\rangle$, $\cos\frac{\alpha_2}{2}|K\rangle\otimes|\uparrow\rangle - \sin\frac{\alpha_2}{2}|K'\rangle\otimes|\downarrow\rangle$

Some of the angles (such as $\alpha, \theta$) have not been specified because they either have very complicated expressions or are evaluated numerically. We have also used the notation,

$$|\pm e_x\rangle = \frac{1}{\sqrt{2}}(|K\rangle \pm |K'\rangle), \quad |\searrow_1\rangle = \begin{pmatrix}\cos\frac{\theta_1}{2}\\ \sin\frac{\theta_1}{2}\end{pmatrix}, \quad |\nearrow_2\rangle = \begin{pmatrix}\cos\frac{\theta_2}{2}\\ -\sin\frac{\theta_2}{2}\end{pmatrix}. \tag{S13}$$

## SIII. TRANSPORT GAPS OF FRACTIONAL QUANTUM HALL STATES IN GRAPHENE

Once the ground state is known for a given set of couplings, the transport gaps $\Delta_{i\to j}$ may be found by considering the energy of creating a quasihole-quasiparticle pair in the limit of large wavenumber. For FQH states with filling $(1, \nu_F)$, the possible excitations are,

- $\Delta_{2\to 3(4)}$: gap of the excited state containing a quasi-hole in $|f_2\rangle$ and a quasi-particle in $|f_{3(4)}\rangle$;
- $\Delta_{1\to 2}$: gap of the excited state containing a quasi-hole in $|f_1\rangle$ and a quasi-particle in $|f_2\rangle$,
- $\Delta_{2\to 2}$: gap of the excited state containing a quasi-hole and a quasi-particle in $|f_2\rangle$.

Out of these, there is one flavor-conserving (f.c.) gap $\Delta_{2\to 2}$, and three flavor-flipping (f.f.) gaps $\Delta_{2\to 3}$, $\Delta_{2\to 4}$, $\Delta_{1\to 2}$.

Since $\Delta_{1\to 2}$ involves charge transfer between fully filled ($|f_1\rangle$) and fractionally filled ($|f_2\rangle$) levels, we may simplify its computation by invoking the invariance of the Hamiltonian under the particle-hole transformation accompanied by an interchange of valleys ($K \rightleftharpoons K'$) and spins ($\uparrow\rightleftharpoons\downarrow$), i.e., $\hat{c}^\dagger_{\tau,\sigma}\to\hat{c}_{\bar\tau,\bar\sigma}$, where $\bar\tau, \bar\sigma$ represent the opposite valley and spin respectively. The inversion of valley and spin is required in order to leave the 1-body terms in the Hamiltonian invariant under the particle-hole transformation. Furthermore, under this transformation, the state at filling $(1,\nu_F)$ is related to that at filling $(1,1,1-\nu_F)$ as shown below,

$$\begin{matrix}|f_1\rangle & |f_2\rangle & |f_3\rangle & |f_4\rangle\\ (\,1 & \nu_F & 0 & 0\,)\end{matrix} \quad\xrightleftharpoons{\text{P.H.}}\quad \begin{matrix}|\bar f_1\rangle & |\bar f_2\rangle & |\bar f_3\rangle & |\bar f_4\rangle\\ (\,0 & 1-\nu_F & 1 & 1\,)\end{matrix}$$

where $|\bar f_i\rangle$'s are related to $|f_i\rangle$ by interchange of spins and valleys. The excited state corresponding to $\Delta_{1\to 2}$ transforms to a state with a quasi-hole in $|\bar f_2\rangle$ and a quasi-particle in $|\bar f_1\rangle$. Since this new excited state involves transfer of charge from a fractionally filled level to an empty level, its energy gap may be evaluated using the framework described below. The invariance of the Hamiltonian under this transformation ensures that $\Delta_{\bar 2\to \bar 1}$ must be identical to $\Delta_{1\to 2}$.

Now we come to the details of the calculation. We first consider the flavor-flipping gaps involving empty levels, such as $\Delta_{2\to 3}$, which is the sum of three separate contributions,

$$\Delta_{2\to 3} = \Delta^{\text{f.f.}}_{C,\nu_F} + \Delta_{1b} + \Delta_F. \tag{S14}$$

The first term $\Delta^{\text{f.f.}}_{C,\nu_F}$ is the Coulomb energy cost of creating a flavor-flipped excitation (in the large wavenumber limit) in the one-component FQH state $|\nu_F\rangle$. The value of $\Delta^{\text{f.f.}}_{C,\nu_F}$ is obtained by exact diagonalization of the Coulomb

| $\Delta_C$ LL | $\Delta^{\text{f.f.}}_{C,\frac{1}{3}}$ | $\Delta^{\text{f.c.}}_{C,\frac{1}{3}}$ | $\Delta^{\text{f.f.}}_{C,\frac{2}{3}}$ | $\Delta^{\text{f.c.}}_{C,\frac{2}{3}}$ |
|---|---|---|---|---|
| **n=0** | 0.075 | 0.10 | 0.067 | 0.10 |
| **n=1** | 0.10 | 0.12 | 0.10 | 0.12 |

TABLE S1: Coulomb contribution [in unit of $e^2/(\kappa\ell)$] to the flavor-flipping (f.f.) gaps and flavor-conserving (f.c.) gaps for $\nu_F=1/3$ and $2/3$.

interaction in finite systems in the spherical geometry [4], with the thermodynamic limit extrapolated (see Table S1). $\Delta_{1b}$ includes the energy cost due to anisotropic interactions between the completely occupied levels and $|f_{2,3}\rangle$, as well as the change in the Zeeman energy due to the transfer of charge from $|f_2\rangle$ to $|f_3\rangle$. Altogether we have, $\Delta_{1b} = \langle f_3|H_{1b}(P_I)|f_3\rangle - \langle f_2|H_{1b}(P_I)|f_2\rangle$, where $H_{1b}(P_I)$ is the Hartree-Fock Hamiltonian constructed out of the projector to the fully filled spinors and the external fields [7],

$$H_{1b}(P_I) = \sum_{a=x,y,z}\left[u_{a,H}\mathrm{Tr}(P_I\tau^a)\tau^a - u_{a,F}\tau^a P_I \tau^a\right] - E_Z\sigma^z, \quad P_I = |f_1\rangle\langle f_1|. \tag{S15}$$

$$u_{a,\mathbf{n},H} = 2\sum_m u_{a,\mathbf{n}}^{(m)}, \quad u_{a,\mathbf{n},F} = 2\sum_m (-1)^m u_{a,\mathbf{n}}^{(m)}. \tag{S16}$$

Note that the factor of 2 in the Hartree and Fock couplings comes due to the relation (S4). The last contribution $\Delta_F$ in Eq. S14 is the energy cost of creating the excitation due to the anisotropic interactions between $|f_2\rangle$ and $|f_3\rangle$. This term vanishes if only the $m=0$ pseudopotential is present in the Hamiltonian [7]. However, this contribution is finite for $m>0$ interactions, which are naturally present in the $\mathbf{n}=1$ LL. To evaluate $\Delta_F$, we need to project the anisotropic interactions to the subspace spanned by the spinors $|f_2\rangle \equiv |\Uparrow\rangle$ & $|f_3\rangle \equiv |\Downarrow\rangle$. We first consider the projection of the matrices $\tau^a$,

$$\tau^a \rightarrow \begin{pmatrix} \langle\Uparrow|\tau^a|\Uparrow\rangle & \langle\Uparrow|\tau^a|\Downarrow\rangle \\ \langle\Downarrow|\tau^a|\Uparrow\rangle & \langle\Downarrow|\tau^a|\Downarrow\rangle \end{pmatrix} \equiv T^a = \sum_b \mathcal{R}_b^a \mathcal{S}^b. \tag{S17}$$

Here, $\mathcal{S}^b$ are Pauli matrices ($b=0,x,y,z$) in the subspace of $|f_2\rangle$ and $|f_3\rangle$, and $\mathcal{R}_b^a$ are the corresponding coefficients, which are constrained to be real by the hermiticity of the projected density matrix. Then the projected anisotropic interaction Eq. (S3) may be written as,

$$(\hat{H}_{\mathbf{n}}^{an})_{\mathrm{projected}} = \frac{1}{2}\sum_{m=0}\sum_a u_{a,\mathbf{n}}^{(m)}\mathbf{U}_{m_1m_2m_3m_4}^{(m)} : \hat{T}_{m_1m_4}^a \hat{T}_{m_2m_3}^a :, \quad \hat{T}_{m_1m_4}^a = T_{\alpha\lambda}^a \hat{c}_{m_1,\alpha}^\dagger \hat{c}_{m_4,\lambda}. \tag{S18}$$

Then the energy gap $\Delta_F$ is, $\Delta_F = \langle \nu_F^{\mathrm{f.f.}}|(\hat{H}_{\mathbf{n}}^{an})_{\mathrm{projected}}|\nu_F^{\mathrm{f.f.}}\rangle - \langle \nu_F|(\hat{H}_{\mathbf{n}}^{an})_{\mathrm{projected}}|\nu_F\rangle$, where $|\nu_F\rangle$ is the FQH state at filling factor $\nu_F$ and $|\nu_F^{\mathrm{f.f.}}\rangle$ is the flavor-flipped excited state on top of it. This expression can be simplified further by noting that both $|\nu_F\rangle$ and $|\nu_F^{\mathrm{f.f.}}\rangle$ are eigenstates of total $\mathcal{S}_z$. We emphasize that this does not mean ordinary spin, but rather refers to matrix in the space $\Uparrow,\Downarrow$. Therefore, only those components of $(\hat{H}_{\mathbf{n}}^{an})_{\mathrm{projected}}$ that conserve the total spin will contribute to the energy. There are two such terms which we consider one-by-one below for the excited state matrix element. The ground state contribution will have the same form. Consider first the diagonal part of the $T$ matrices: $\mathcal{R}_0^a \mathcal{S}_0 + \mathcal{R}_z^a \mathcal{S}_z = \mathrm{diag}(\mathcal{R}_\Uparrow^a, \mathcal{R}_\Downarrow^a)$. The only $\mathcal{S}_z$ conserving combination of these is,

$$\frac{1}{2}\sum_a\sum_m u_{a,\mathbf{n}}^{(m)}\mathbf{U}_{m_1m_2m_3m_4}^{(m)} \times \langle\nu_F^{\mathrm{f.f.}}|:\left(\mathcal{R}_\Uparrow^a \hat{c}_{m_1,\Uparrow}^\dagger \hat{c}_{m_4,\Uparrow} + \mathcal{R}_\Downarrow^a \hat{c}_{m_1,\Downarrow}^\dagger \hat{c}_{m_4,\Downarrow}\right)\left(\mathcal{R}_\Uparrow^a \hat{c}_{m_2,\Uparrow}^\dagger \hat{c}_{m_3,\Uparrow} + \mathcal{R}_\Downarrow^a \hat{c}_{m_2,\Downarrow}^\dagger \hat{c}_{m_3,\Downarrow}\right):|\nu_F^{\mathrm{f.f.}}\rangle$$

$$= \frac{1}{2}\sum_a\sum_m u_{a,\mathbf{n}}^{(m)}\Bigg[(\mathcal{R}_\Uparrow^a)^2 \underbrace{\langle\nu_F^{\mathrm{f.f.}}|\mathbf{U}_{m_1m_2m_3m_4}^{(m)}:\hat{c}_{m_1,\Uparrow}^\dagger \hat{c}_{m_4,\Uparrow}\hat{c}_{m_2,\Uparrow}^\dagger \hat{c}_{m_3,\Uparrow}:|\nu_F^{\mathrm{f.f.}}\rangle}_{\equiv\langle\hat{\mathbf{U}}_{\Uparrow\Uparrow}^{(m)}\rangle_{\nu_F}^{\mathrm{f.f.}}}$$

$$+ \mathcal{R}_\Uparrow^a \mathcal{R}_\Downarrow^a \underbrace{\langle\nu_F^{\mathrm{f.f.}}|2\,\mathbf{U}_{m_1m_2m_3m_4}^{(m)}:\hat{c}_{m_1,\Uparrow}^\dagger \hat{c}_{m_4,\Uparrow}\hat{c}_{m_2,\Downarrow}^\dagger \hat{c}_{m_3,\Downarrow}:|\nu_F^{\mathrm{f.f.}}\rangle}_{\equiv\langle\hat{\mathbf{U}}_{\Uparrow\Downarrow}^{(m)}\rangle_{\nu_F}^{\mathrm{f.f.}}}$$

$$+ (\mathcal{R}_\Downarrow^a)^2 \underbrace{\langle\nu_F^{\mathrm{f.f.}}|\mathbf{U}_{m_1m_2m_3m_4}^{(m)}:\hat{c}_{m_1,\Downarrow}^\dagger \hat{c}_{m_4,\Downarrow}\hat{c}_{m_2,\Downarrow}^\dagger \hat{c}_{m_3,\Downarrow}:|\nu_F^{\mathrm{f.f.}}\rangle}_{\equiv\langle\hat{\mathbf{U}}_{\Downarrow\Downarrow}^{(m)}\rangle_{\nu_F}^{\mathrm{f.f.}}}\Bigg]. \tag{S19}$$

| $m$ | $\delta\langle\hat{\mathbf{U}}^{(m)}_{\Uparrow\Uparrow}\rangle_{\frac{1}{3}}$ | $\delta\langle\hat{\mathbf{U}}^{(m)}_{\Uparrow\Downarrow}\rangle_{\frac{1}{3}}$ | $\delta\langle\hat{\mathbf{U}}^{(m)}_{\Downarrow\Downarrow}\rangle_{\frac{1}{3}}$ | $\delta\langle\hat{\mathbf{U}}^{(m)}_{\Uparrow\Uparrow}\rangle_{\frac{2}{3}}$ | $\delta\langle\hat{\mathbf{U}}^{(m)}_{\Uparrow\Downarrow}\rangle_{\frac{2}{3}}$ | $\delta\langle\hat{\mathbf{U}}^{(m)}_{\Downarrow\Downarrow}\rangle_{\frac{2}{3}}$ |
|---|---|---|---|---|---|---|
| 1 | 0.06(1) | $\sim 0$ | 0 | -0.32(1) | $\sim 0$ | 0 |
| 2 | 0 | 0.93(1) | 0 | 0 | 2.12(1) | 0 |

TABLE S2: Change in the average pair amplitudes of $\nu = \frac{1}{3}$ & $\frac{2}{3}$ FQHE state due to 1 electron flip for relative angular momentum $m$=0, 1, 2. Notes: (1) Since only 1 "spin" flips from $\Uparrow$ to $\Downarrow$, $\delta\langle\hat{\mathbf{U}}^{(m)}_{\Downarrow\Downarrow}\rangle_{\nu_F}$ is exactly zero for any $m$; (2) For odd $m$, $\delta\langle\hat{\mathbf{U}}^{(1)}_{\Uparrow\Downarrow}\rangle_{\nu_F}$ is $\sim \mathcal{O}(10^{-3})$ and hence we neglect it in our analysis; (3) For even $m$, fermionic antisymmetry will force $\langle\hat{\mathbf{U}}^{(m)}_{\Uparrow\Uparrow}\rangle_{\nu_F} = \langle\hat{\mathbf{U}}^{(m)}_{\Uparrow\Uparrow}\rangle^{\text{f.f.}}_{\nu_F} = 0$.

Next, consider the non-diagonal matrices $\mathcal{S}^x$ & $\mathcal{S}^y$. The "spin"-conserving combination involving these is,

$$\frac{1}{2}\sum_a \sum_m u^{(m)}_{a,\mathbf{n}} \mathbf{U}^{(m)}_{m_1 m_2 m_3 m_4} \langle\nu^{\text{f.f.}}_F| : \left(\mathcal{R}^a_x \hat{S}^x + \mathcal{R}^a_y \hat{S}^y\right)_{m_1 m_4} \left(\mathcal{R}^a_x \hat{S}^x + \mathcal{R}^a_y \hat{S}^y\right)_{m_2 m_3} : |\nu^{\text{f.f.}}_F\rangle$$

$$= \frac{1}{2}\sum_a \sum_m u^{(m)}_{a,\mathbf{n}} \mathbf{U}^{(m)}_{m_1 m_2 m_3 m_4} \langle\nu^{\text{f.f.}}_F| : \left[\mathcal{R}^a_x\left(\hat{c}^\dagger_{m_1,\Uparrow}\hat{c}_{m_4,\Downarrow} + \hat{c}^\dagger_{m_1,\Downarrow}\hat{c}_{m_4,\Uparrow}\right) - i\mathcal{R}^a_y\left(\hat{c}^\dagger_{m_1,\Uparrow}\hat{c}_{m_4,\Downarrow} - \hat{c}^\dagger_{m_1,\Downarrow}\hat{c}_{m_4,\Uparrow}\right)\right]$$

$$\times \left[\mathcal{R}^a_x\left(\hat{c}^\dagger_{m_2,\Uparrow}\hat{c}_{m_3,\Downarrow} + \hat{c}^\dagger_{m_2,\Downarrow}\hat{c}_{m_3,\Uparrow}\right) - i\mathcal{R}^a_y\left(\hat{c}^\dagger_{m_2,\Uparrow}\hat{c}_{m_3,\Downarrow} - \hat{c}^\dagger_{m_2,\Downarrow}\hat{c}_{m_3,\Uparrow}\right)\right] : |\nu^{\text{f.f.}}_F\rangle$$

$$= \frac{1}{2}\sum_a \sum_m u^{(m)}_{a,\mathbf{n}} \left(\mathcal{R}^a_x \mathcal{R}^a_x + \mathcal{R}^a_y \mathcal{R}^a_y\right) \times (-1)^{m+1} \underbrace{\langle\nu^{\text{f.f.}}_F|2\mathbf{U}^{(m)}_{m_1 m_2 m_3 m_4} : \hat{c}^\dagger_{m_1,\Uparrow}\hat{c}_{m_4,\Uparrow}\hat{c}^\dagger_{m_2,\Downarrow}\hat{c}_{m_3,\Downarrow} : |\nu^{\text{f.f.}}_F\rangle}_{\equiv \langle\mathbf{U}^{(m)}_{\Uparrow\Downarrow}\rangle^{\text{f.f.}}_{\nu_F}}. \quad \text{(S20)}$$

Clearly, the expressions for the ground state matrix element will have the same form as (S19, S20) except that the pair amplitudes would correspond to the ground state. Then, combining everything together, we have,

$$\Delta_F = \frac{1}{2}\sum_{a,m} u^{(m)}_{a,\mathbf{n}} \left[\left(\mathcal{R}^a_\Uparrow\right)^2 \delta\langle\hat{\mathbf{U}}^{(m)}_{\Uparrow\Uparrow}\rangle_{\nu_F} + \left(\mathcal{R}^a_\Uparrow \mathcal{R}^a_\Downarrow - (-1)^m\left(\mathcal{R}^a_x \mathcal{R}^a_x + \mathcal{R}^a_y \mathcal{R}^a_y\right)\right)\delta\langle\hat{\mathbf{U}}^{(m)}_{\Uparrow\Downarrow}\rangle_{\nu_F} + \left(\mathcal{R}^a_\Downarrow\right)^2 \delta\langle\hat{\mathbf{U}}^{(m)}_{\Downarrow\Downarrow}\rangle_{\nu_F}\right]. \quad \text{(S21)}$$

Here, $\delta\langle\hat{\mathbf{U}}^{(m)}_{b_1 b_2}\rangle^{\text{f.f.}}_{\nu_F} = \langle\hat{\mathbf{U}}^{(m)}_{b_1 b_2}\rangle^{\text{f.f.}}_{\nu_F} - \langle\hat{\mathbf{U}}^{(m)}_{b_1 b_2}\rangle_{\nu_F}$ is the difference of the pair amplitudes in the excited and ground states ($b_{1,2} = \Uparrow, \Downarrow$). Note that for 1-component FQH states only $\langle\hat{\mathbf{U}}^{(m)}_{\Uparrow\Uparrow}\rangle_{\nu_F}$ is finite. The pair amplitudes in both ground and excited states may be evaluated via exact diagonalization (The difference is presented in Table S2).

The other flavor flipping gaps $\Delta_{2\to 4}$ and $\Delta_{1\to 2}$ may be evaluated similarly. For the latter, we consider the particle-hole conjugate, as described earlier, and evaluate the gap $\Delta_{\bar{2}\to\bar{1}}$ with the convention $|\Uparrow\rangle = |\bar{f}_2\rangle$, $|\Downarrow\rangle = |\bar{f}_1\rangle$ and $P_I = |\bar{f}_3\rangle\langle\bar{f}_3| + |\bar{f}_4\rangle\langle\bar{f}_4|$.

The flavor conserving gap $\Delta_{2\to 2}$ is the sum of two contributions, $\Delta_{2\to 2} = \Delta^{\text{f.c.}}_{C,\nu_F} + \Delta_F$. $\Delta^{\text{f.c.}}_{C,\nu_F}$ is the Coulomb charge gap of the flavor-conserving excitation (evaluated through exact diagonalization; see Table S1). $\Delta_F$ is the charge gap due to the anisotropic interaction within the $|f_2\rangle$ LL. As before, it may be expressed as,

$$\Delta_F = \langle\nu^{\text{f.c.}}_F|\hat{H}^{an}_\mathbf{n}|\nu^{\text{f.c.}}_F\rangle - \langle\nu_F|\hat{H}^{an}_\mathbf{n}|\nu_F\rangle$$

$$= \frac{1}{2}\sum_{\text{odd } m}\sum_a u^{(m)}_{a,\mathbf{n}} \tau^a_{22}\tau^a_{22}\left(\langle\nu^{\text{f.c.}}_F|\mathbf{U}^{(m)}_{m_1 m_2 m_3 m_4}\hat{c}^\dagger_{m_1,2}\hat{c}^\dagger_{m_2,2}\hat{c}_{m_3,2}\hat{c}_{m_4,2}|\nu^{\text{f.c.}}_F\rangle - \langle\nu_F|\mathbf{U}^{(m)}_{m_1 m_2 m_3 m_4}\hat{c}^\dagger_{m_1,2}\hat{c}^\dagger_{m_2,2}\hat{c}_{m_3,2}\hat{c}_{m_4,2}|\nu_F\rangle\right)$$

$$= \frac{1}{2}\sum_a \left[\sum_{\text{odd } m} \underbrace{\left(\langle\nu^{\text{f.c.}}_F|\hat{\mathbf{U}}^{(m)}|\nu^{\text{f.c.}}_F\rangle - \langle\nu_F|\hat{\mathbf{U}}^{(m)}|\nu_F\rangle\right)}_{\delta\langle\hat{\mathbf{U}}^{(m)}_{\text{f.c.}}\rangle_{\nu_F}} u^{(m)}_{a,\mathbf{n}}\right] \times \frac{1}{2}\left[\text{Tr}^2(P_2\tau^a) + \text{Tr}(P_2\tau^a P_2\tau^a)\right]. \quad \text{(S22)}$$

Here, we have used $\tau^a_{22}\tau^a_{22} = \langle f_2|\tau^a|f_2\rangle^2 = \text{Tr}(P_2\tau^a)^2 = \frac{1}{2}\left[\text{Tr}^2(P_2\tau^a) + \text{Tr}(P_2\tau^a P_2\tau^a)\right]$. In the $\mathbf{n} = 0$ LL, this term vanishes since the $m = 0$ pseudopotential does not contribute to this gap. In the $\mathbf{n} = 1$ LL, the $m = 1$ pair amplitude enters the computation. Exact diagonalization on finite systems extrapolated to the thermodynamic limit gives the numerical values $\delta\langle\hat{\mathbf{U}}^{(m)}_{\text{f.c.}}\rangle_{\frac{1}{3}} = \delta\langle\hat{\mathbf{U}}^{(m)}_{\text{f.c.}}\rangle_{\frac{2}{3}} = 0.46(1)$.

## SIV.  THE ZEEMAN DEPENDENCE OF THE GAPS IN THE V/SAFM AND VPAFM PHASES

In the 1LLs we have assumed $g_{z0} = g_{\perp 0} = 0$. At this point, in the second quadrant of the $g_{\perp\perp} - g_{z\perp}$ plane, the VPAFM and the VSAFM phases are degenerate. At first glance, it appears that both the V/SAFM and VPAFM phases are ideal for obtaining a decreasing transport gap as $E_Z$ increases, as shown by the following simple argument. The occupied spinors in the V/SAFM phase are $|f_1\rangle = |K \uparrow\rangle$ and $|f_2\rangle = |K' \downarrow\rangle$. Taking a quasiparticle from the fractionally occupied $|K' \downarrow\rangle$ to any empty level while flipping the spin will give a contribution to the gap of $-2E_Z$. Similarly, the occupied spinors in the VPAFM phase are $|f_1\rangle = |K \uparrow\rangle$ and $|f_2\rangle = |K \downarrow\rangle$. The same argument as above applies to this state as well.

Unfortunately, this expectation does not work quantitatively due to other contributions to the gaps. Both the V/SAFM and VPAFM states are simple enough that all the gaps can be computed analytically. Here are the anisotropic and Zeeman contributions to the gaps for the V/SAFM state (we have dropped $\Delta_F$ in the following expressions as that is very small in this case):

$$\Delta_{2\to 3} = -2u_{\perp F} - 2E_Z + 0.1\frac{e^2}{\kappa\ell}$$
$$\Delta_{2\to 4} = 2u_{zH} + 0.1\frac{e^2}{\kappa\ell}$$
$$\Delta_{1\to 2} = 2E_Z + 0.1\frac{e^2}{\kappa\ell}$$
$$\Delta_{2\to 2} = 0.12\frac{e^2}{\kappa\ell}. \tag{S23}$$

Since $u_{\perp F}$ has to be negative for the system to be in the V/SAFM phase, the gap that decreases with $E_Z$ is not the lowest gap in the experimental range of $E_Z$. Similarly, the gaps for the VPAFM phase are (without $\Delta_F$ since it is small):

$$\Delta_{2\to 3} = -2u_{\perp F} - 2u_{zH} - 2E_Z + 0.1\frac{e^2}{\kappa\ell}$$
$$\Delta_{2\to 4} = -2u_{zH} + 0.1\frac{e^2}{\kappa\ell}$$
$$\Delta_{1\to 2} = 2E_Z + 0.1\frac{e^2}{\kappa\ell}$$
$$\Delta_{2\to 2} = 0.12\frac{e^2}{\kappa\ell}. \tag{S24}$$

Given our assumption that $g_{z0} = g_{\perp 0} = 0$, we obtain $u_{zH} = 0$ and the gaps are identical to those in the V/SAFM phase.

---

[1] F. Amet, A. J. Bestwick, J. R. Williams, L. Balicas, K. Watanabe, T. Taniguchi, and D. Goldhaber-Gordon, Composite fermions and broken symmetries in graphene, Nat. Commun. **6**, 5838 (2015).

[2] I. L. Aleiner, D. E. Kharzeev, and A. M. Tsvelik, Spontaneous symmetry breaking in graphene subjected to an in-plane magnetic field, Phys. Rev. B **76**, 195415 (2007).

[3] J. An, A. C. Balram, and G. Murthy, Magnetic and lattice ordered fractional quantum Hall phases in graphene, Phys. Rev. B **110**, L081103 (2024).

[4] F. D. M. Haldane, Fractional quantization of the Hall effect: A hierarchy of incompressible quantum fluid states, Phys. Rev. Lett. **51**, 605 (1983).

[5] J. An, A. C. Balram, U. Khanna, and G. Murthy, Fractional quantum Hall coexistence phases in higher Landau levels of graphene, Phys. Rev. B **111**, 045110 (2025).

[6] Y. Lian and M. O. Goerbig, Spin-valley skyrmions in graphene at filling factor $\nu = -1$, Phys. Rev. B **95**, 245428 (2017).

[7] I. Sodemann and A. H. MacDonald, Broken SU(4) symmetry and the fractional quantum Hall effect in graphene, Phys. Rev. Lett. **112**, 126804 (2014).



\end{document}